\documentclass[5p,times]{elsarticle}

\usepackage{microtype}

\usepackage{multirow}

\usepackage{caption}
\usepackage{subcaption}

\usepackage{url}

\usepackage{amssymb}
\usepackage{amsthm}

\usepackage[switch]{lineno}

\usepackage{enumitem}
\setlist[itemize]{align=parleft,left=0pt..1em}

\usepackage{soul}
\usepackage[dvipsnames]{xcolor}

\usepackage[final]{changes}

\journal{Future Generation Computer Systems}

\begin{document}

\fbox{\begin{minipage}[b][1cm][c]{18cm}
\footnotesize This article has been accepted for publication in the Future Generation Computer Systems, Elsevier. This is the author's version which has not been fully edited and content may change prior to final publication. Citation information: \url{https://doi.org/10.1016/j.future.2022.05.027}
\end{minipage}}

\begin{frontmatter}



\title{Towards End-to-End Application Slicing in Multi-access Edge Computing systems: Architecture Discussion and Proof-of-Concept}


\author[cnr]{Simone Bolettieri}

\affiliation[cnr]{organization={Institute of Informatics and Telematics (IIT) - Italian National Research Council (CNR)},
            addressline={Via G. Moruzzi 1}, 
            city={Pisa},
            postcode={56124}, 
            country={Italy}}

\author[nokia]{Dinh Thai Bui}
\author[cnr]{Raffaele Bruno}

\affiliation[nokia]{organization={Nokia Bell Labs},
            addressline={1 route de Villejust}, 
            city={Nozay},
            postcode={91620}, 
            country={France}}

\begin{abstract}
Network slicing is one of the most critical 5G pillars. It allows for sharing a 5G infrastructure among different tenants leading to improved service customisation and increased operators' revenues. Concurrently, introducing the Multi-access Edge Computing (MEC) into 5G to support time-critical applications raises the need to integrate this distributed computing infrastructure to the 5G network slicing framework. Indeed, end-to-end latency guarantees require the end-to-end management of slice resources. For this purpose, after discussing the main gaps in the state-of-the-art with regards to such an objective, we propose a novel slicing architecture that enables the management and orchestration of slice segments that span over all the domains of an end-to-end application service, including the MEC. We also show how this general management architecture can be instantiated into a multi-tenant MEC infrastructure. A preliminary implementation of the proposed architecture focusing on the MEC domain is also provided, together with performance tests to validate the feasibility and efficacy of our design approach.
\end{abstract}




\end{frontmatter}


\section{Introduction}
\label{sec:introduction}
\noindent
Amongst different design focuses, 5G standards have specifically addressed low-latency requirements to  cater safety-critical and mission-critical applications~\cite{2020_comst_5G_latency}. Industrial automation within Industry 4.0 is an example of new perspectives that are triggered by such a new capability. To reach low-latency targets, the 5G New Radio (NR) standard has defined the Ultra-Reliable Low Latency Communication (URLLC), which allows for reaching a few ms one-way latency with small radio frames~\cite{2021_access_urllc}. Furthermore, to improve the end-to-end latency, the 5G architecture has integrated distributed cloud infrastructures enabled by the ETSI standard for Multi-access Edge Computing (MEC)~\cite{MEC003}, which brings compute, storage and network resources next to the end-user devices. Indeed, processing data at the network’s edge instead of transferring it to remote cloud data centres is beneficial not only to reduce bandwidth requirements and access times but also to increase the ability of the network providers to satisfy the needs of complex real-time applications~\cite{2018_comst_mec_iot,2021_jnca_edge}.  

Another important pillar of the 5G standards is the network slice concept that allows sharing the same 5G infrastructure amongst multiple service providers. This aims at improving network flexibility and service customisation and, in the end, maximising 5G operator revenue. There are several definition of a network slice~\cite{NFV028,NFV012,NGMN028,3GPPTR28801,IETFFarrel}. However, all have agreed that a (network) slice is a logical network composed of dedicated and shared network resources providing specific network capabilities and characteristics. One of the key features of network slicing is isolation between (network) slice instances at different levels, namely at the data plane, the control plane and the management plane. Furthermore, a network slice needs to span not only over different architectural levels but also over different technological domains (e.g., access, core and transport networks). Therefore, the definition of end-to-end (E2E) network slicing frameworks that seamlessly interconnect different slice segments (or slice subnets in 5G terminology) is an open research challenge~\cite{ZSM003}. Furthermore, as latency should be managed end-to-end, there is a clear need to establish and manage an E2E network slice including the MEC~\cite{MEC024}. The most mature proposal to enable the instantiation of a network slice embedding MEC components is the MEC-in-NFV architecture~\cite{MEC003}, which proposes to use Network Function Virtualisation (NFV) technologies to virtualise MEC applications and MEC management entities. Recently, a few research works have attempted to extend the MEC-in-NFV orchestration architecture to allow not only multiple tenants (namely MEC applications) to co-exist in the same network slice but also the sharing of multiple network slices with the same tenant~\cite{Cominardi2020,2020_MNET_MEC_subslice}. However, those proposals implicitly assume that only an end customer (e.g., a UE requesting a specific service) can be a tenant of a MEC platform. On the contrary, in this study, we advocate a more articulated multi-tenancy model in which an operator of a MEC infrastructure (we call it MEC Owner) interacts with the 5G system and offers virtualised MEC environments to multiple tenants (we call them MEC Customers). Then, a MEC Customer acts as an application service provider towards its end users. This architectural design has multiple benefits. First, it allows the deployment of MEC environments that are dedicated (and optimised) to specific application domains. For instance, a MEC Customer can offer specialised services to car manufacturers that want to deploy automotive applications. In contrast, another MEC Customer can offer specialised services to law enforcement agencies in a city that want to deploy video analytics applications (e.g. traffic monitoring, crowd control). However, both MEC Customers can share the same MEC infrastructure deployed by a MEC Owner in a given city. Second, this architecture facilitates a clear distinction between management roles and businesses responsibilities of different stakeholders, as the orchestration of edge-related resources is clearly separated from the orchestration of network-related resources. We argue that this is an important improvement over existing proposals that normally delegate the management and orchestration of application components to the orchestrator already contained within an NFV architecture, as in~\cite{Cominardi2020}. To some extent, this is equivalent to moving from a sliced MEC management platform to a fully \textit{sliced MEC orchestration} capability. Finally, our proposal also permits to clearly distinguish responsibilities in terms of contribution to application latency budget, as the MEC Owner is essentially responsible for the network latency in the edge platform, while the MEC Customer is essentially responsible for data processing and storage latencies.

This paper achieves the above goals through a fourfold contribution. First, we introduce the new concepts of Application Slice (APS) and Application Service (AS), which allows for better distinguishing between the management and orchestration of application and network services. Second, we propose an end-to-end slice architecture together with the related operational and business roles, which spans over all the domains of an end-to-end application service. This architecture design allows for considering all processing and storage latencies, especially those within the MEC. Third, we propose an overall management architecture that allows the life-cycle management of such end-to-end slices, following two distinct deployment models. We also show how this general management architecture can be instantiated into a multi-tenant MEC infrastructure with two layers of virtualisation and orchestration, which is enabled by emerging nested virtualisation capabilities~\cite{MEC024}. Fourth, we propose a preliminary implementation of the described architecture focusing on the MEC domain and the MEC Customer orchestrator. We also show experimental results to validate the feasibility of our design approach and its ability to support performance isolation between deployed application slices.  

The rest of this paper is organised as follows. Section~\ref{sec:background} provides background information about the network slicing concept, the 3GPP slice management architecture and the MEC standard. Section~\ref{sec:app_slice} presents our proposed end-to-end application slice management architecture and how it can be instantiated in the context of MEC and 5G systems. Section~\ref{sec:implementation} describes our proof-of-concept implementation and Section~\ref{sec:evaluation} shows experimental results. Finally, we conclude with section~\ref{sec:conclusions} where we provide some hints on our future works. 

\section{Background}
\label{sec:background}
\noindent
Network slicing is not a novel idea. For instance, early network designers proposed to set up virtual LANs to provide network segmentation at the data link layer, addressing issues such as scalability, security, and network management of switched Ethernet networks~\cite{2011_commag_vlan}. However, over the years, the concept of network slicing has significantly evolved. One of the main drivers of this technology has become the need for network service flexibility and programmability to efficiently accommodate diverse business use cases~\cite{2018_COMST_slicing_survey}. Nowadays, a network slice is commonly defined as a \textit{logically isolated network over multi-domain, multi-technology physical networks, which provides specific network capabilities and characteristics} to support certain communication services serving a business purpose while ensuring functional and performance isolation~\cite{2020_ACCESS_slicing_survey}. While there is a consolidated high-level view of network slicing, various network slicing models are currently under definition depending on the virtualisation technology, the network architecture, and the standard development organisation (SDO)~\cite{NGMN028,2017_foukas_5G_slicing,Cominardi2020}. The interested reader is referred to~\cite{MEC024} for a comprehensive comparison of different slicing concepts. This section overviews the 3GPP approach to network slicing as it provides the foundation for our proposed E2E slicing management architecture. Furthermore, we discuss current trends to integrate support for network slicing into the MEC standard, and we identify some gaps to be filled to extend slicing to a multi-tenant edge in a common management architecture. For the sake of presentation clarity, Table~\ref{tab:abbrev} summarises the abbreviations we use in this paper. 
\begin{table}[t]
    \centering
    \small
    \begin{tabular}{|p{0.1\textwidth}|p{0.35\textwidth}|}
        \hline
        \bf Notation & \bf Definition \\
        \hline
         ACF    & Application Component Function \\
         APSMF  & Application Slice Management Function \\
         APSS   & Application Slice Subnet \\
         APSSMF & Application Slice Subnet Management Function \\
         APS    & Application Slice \\
         AS     & Application Service \\
         ASMF   & Application Service Management Function \\
          \added{CN} & \added{Core Network} \\
          \added{CNI} & \added{Container Network Interface} \\
         CS     & Communication Service \\
         CSMF   & Communication Service Management Function \\
         E2E    & End-to-End \\
         LCM    & Lifecycle Management \\
         \added{MANO} & \added{Management and Orchestration stack} \\
         MEO    & MEC Orchestrator\\
         MEAO   & MEC Application Orchestrator\\
         MAPSS & MEC Application Slice Subnet \\
         MAPSSD & MEC Application Slice Subnet Descriptor \\
         MECO   & MEC Customer Orchestrator\\
         MECPM  & MEC Customer Platform Manager\\
         MEOO   & MEC Owner Orchestrator\\
         MENSM  & MEC Network Slice Manager\\
         MEPM   & MEC Platform Manager \\
         NF     & Network Function \\
         NSMF   & Network Slice Management Function \\
         NSSMF  & Network Slice Subnet Management Function \\
         NFVI   & NFV Infrastructure \\
         NST    & Network Slice Template \\
         PNF    & Physical Network Function \\
         \added{RAN} & \added{Radio Access Network} \\
         SLA    & Service Level Agreement \\
         SLS    & Service Level Specification \\
         VNF    & Virtual Network Function  \\
         VNFM   & VNF LCM Manager \\
         \hline
    \end{tabular}
    \caption{Summary of notation and abbreviations}
    \label{tab:abbrev}
\end{table}
\subsection{3GPP network slice concept and building blocks}
\label{sec:3gpp_slicing}
\noindent
The 3GPP approach to network slicing is inherited from the NGMN network slicing concept~\cite{NGMN028}, and it is based on the distinction between network slices and network slice instances (NSIs). Specifically, a network slice is defined as "\textit{a logical network that provides specific network capabilities and network characteristics, supporting various service properties for network slice customers}" ~\cite{3GPPTS28530}, while  a NSI is an activated network slice, namely "...\textit{a set of network functions and the resources for these network functions which are arranged and configured ... to meet certain network characteristics}..." that are required by a communication service~\cite{3GPPTR28801}. Following the relationship between CS and NS, the 3GPP network slicing architecture is organised into three distinct logical layers: 1) a service instance layer, 2) a network slice instance layer, and 3) a resource layer~\cite{3GPPTR28801}. The first layer encompasses the service instances (a.k.a. Communication services (CS) in 5G networks) that are associated with service-level agreements (SLA), namely business contracts between the service provider and the clients, which specify the service levels to be ensured. The second layer comprises the NSIs that are deployed to serve the CS requirements. Each NSI is associated to a \textit{Network Slice Template} (NST), also referred to as network slice blueprint~\cite{NGMN028}, which describes the general structure and configuration of the network slice, and the \textit{Service Level Specification} (SLS), which lists the set of technical attributes that have to be satisfied by the deployed network slice. In other words, the SLS translates the business objectives of the SLA to network characteristics. Finally, the third layer includes the necessary physical (hardware and software) and logical resources to support the NSIs.

\begin{figure}[ht]
\centering
\includegraphics[clip,trim= 2cm 2.5cm 1cm 2cm,width=0.5\textwidth]{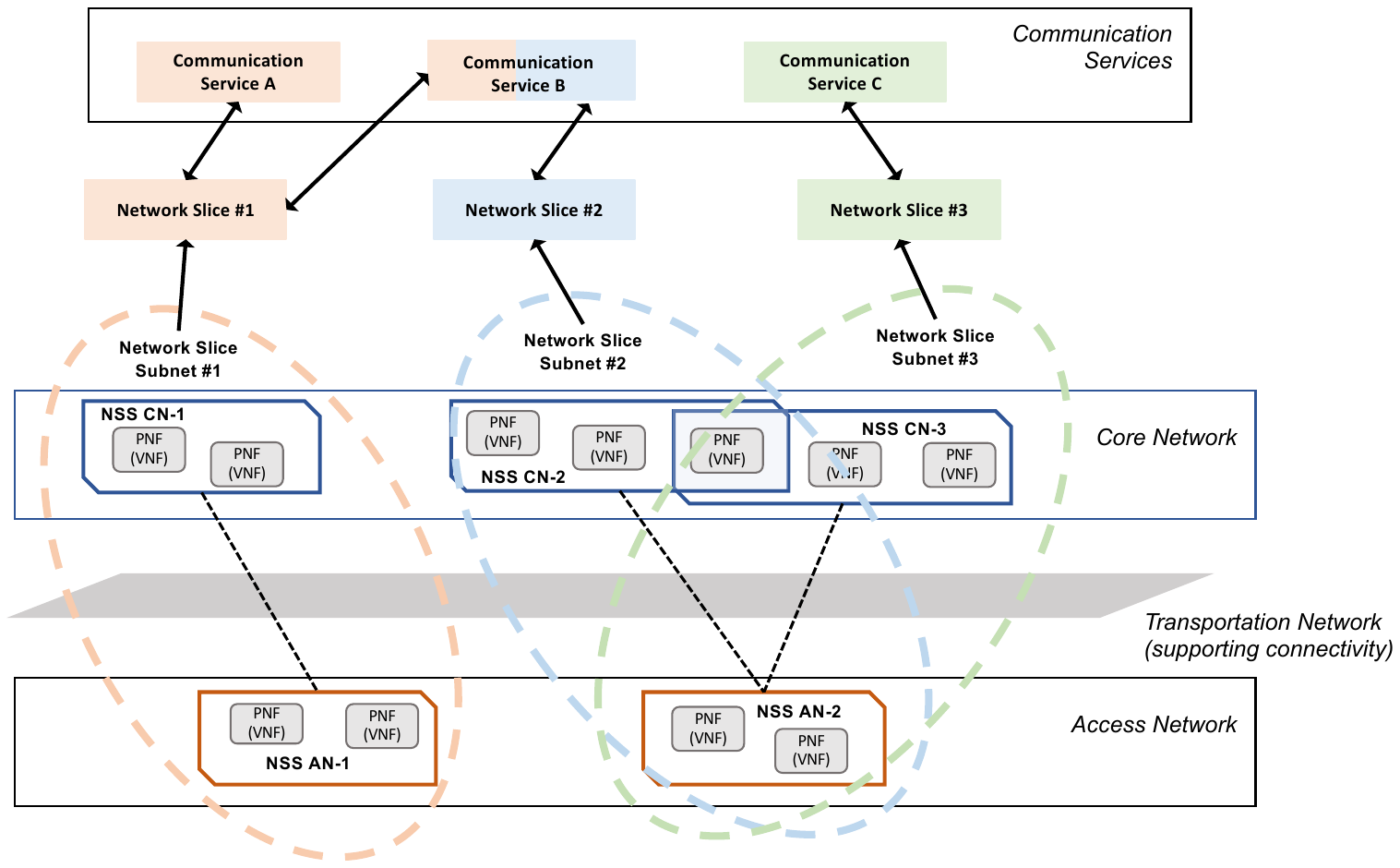}
\caption{Exemplary illustration of the relationships between communication services, network slices, network slice subnets and resources/network functions}
\label{fig:slice_subnet}
\end{figure}

The 3GPP management architecture recognises that a network slice spans different technical domains (or \textit{segment}). Specifically, a network slices spans across the radio access network (RAN), the core network (CN), and the transport network (TN), each segment having separate scope and technologies~\cite{2021_gsma_TR_e2eslicing}. For this reason, the 3GPP network slicing architecture introduces the concept of \textit{Network Slice Subnets} (NSSs), defined as "...\textit{a set of network functions and the associated resources (e.g. compute, storage and networking resources) supporting network slice}." ~\cite{3GPPTS28530}. If a 3GPP CS can be seen as a business wrapper of one or more network slices, in turn, a network slice is a wrapper of one or more network slice subnets with some Service Level Specification (SLS). Figure~\ref{fig:slice_subnet} illustrates the relationship between these different concepts, following an example of~\cite{3GPPTS28530}.
In the figure, there are several points worth noting. Network Functions (NF) refer to processing functions of the 5G network (both access and core networks), which expose APIs to provide one or more services to other NFs, following the producer-consumer concept. NFs include physical network nodes (namely Physical Network Functions or PNF) and Virtual Network Functions (VNF). We remind that a VNF is a software implementation of a network function within an NFV infrastructure (NFVI). In Figure~\ref{fig:slice_subnet}, network slice subnet AN-1 and network slice subnet AN-2 each contain distinct sets of NFs of the RAN, while network slice subnet CN-2 and network slice subnet CN-3 partly share some of the NFs of the core network. The network operator offers network slice subnet 1 as a network slice 1 to CS A. For this purpose, the network operator associates the SLS derived from the SLA of CS A to network slice subnet 1. Note that a network slice can satisfy the service requirements of different communication services (e.g., network slice 1 in Figure~\ref{fig:slice_subnet} is serving both CS A and CS B). Finally, it is helpful to point out that the deployment template of a network slice subnet should contain the descriptors of its constituent NFs and information relevant to the links between these NFs, such as topology of connections and QoS attributes for these links (e.g. bandwidth). The latter information can be represented in the form of a service graph, using the data information model defined by the ETSI NFV standard for describing the Virtualised Network Function Forwarding Graph (VNFFG)~\cite{NFV012}.
\subsection{3GPP network slice management architecture}
\label{subsec:mgmt_architecture}
\begin{figure}[ht]
\centering
\includegraphics[width=0.4\textwidth]{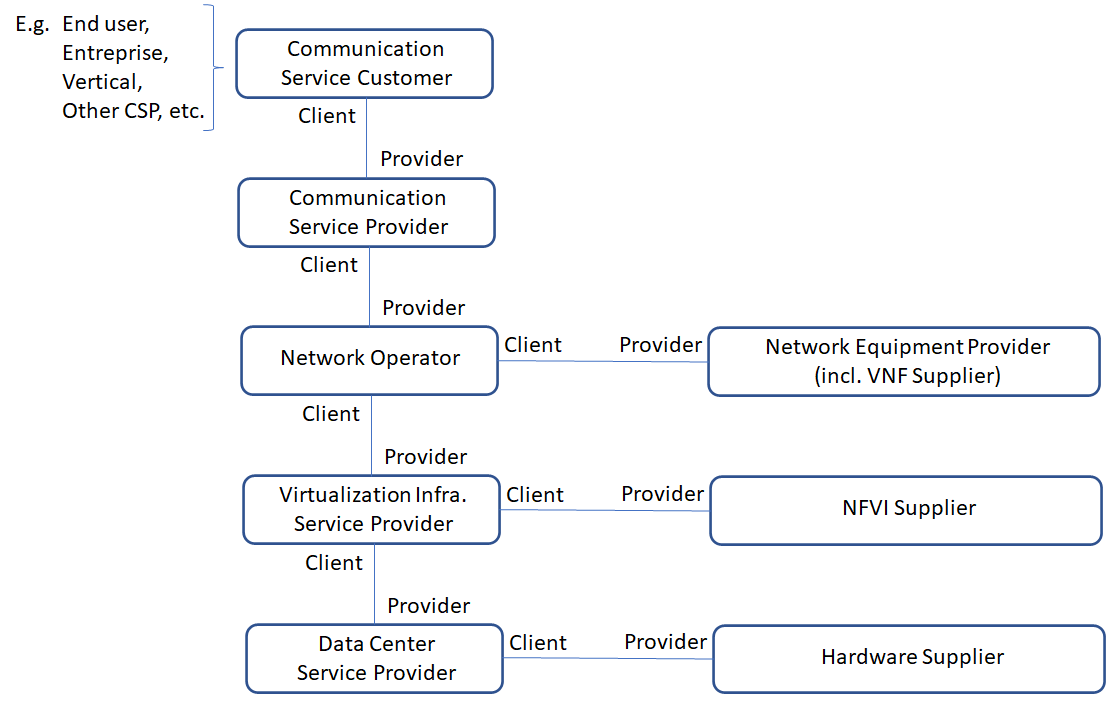}
\caption{High Level Roles in Network Slice Framework}
\label{2_figRoles}
\end{figure}

\noindent
In order to lay the ground for the design of an overall network slice management architecture, the 3GPP has defined high-level operational roles, which permit to draw clear boundaries in terms of operational responsibilities. Figure \ref{2_figRoles} illustrates the different roles as identified by the 3GPP \cite{3GPPTS28530}, which are defined as follows:
\renewcommand\labelitemi{$\bullet$}
\begin{itemize}[noitemsep,topsep=2pt]
    \item Communication Service Customer (CSC): consumes communication services \added{- e.g. end-users or vertical enterprises pay for their communication services.}
    \item Communication Service Provider (CSP): provides communication services that are designed, built and operated with (or without) a network slice \added{- e.g. Virgin Mobile provides public mobile subscriptions, AT\&T provides private network communication services to enterprises.}
    \item Network Operator (NOP): designs, builds and operates network slices \added{- e.g. AT\&T, Verizon.}
    \item Network Equipment Provider (NEP): supplies network equipment including VNFs to a network operator \added{- e.g. Nokia, Cisco, Ericsson.}
    \item Virtualisation Infrastructure Service Provider (VISP): provides virtualised infrastructure services  \added{- e.g. AWS, MS Azure, GCP provide PaaS.}
    \item Data Centre Service Provider (DCSP): provides data centre services \added{- e.g. AWS, MS Azure, GCP.}
    \item NFVI Supplier: supplies a network function virtualisation infrastructure \added{- e.g. VMWARE supplies Hypervisor and related management functions.}
    \item Hardware Supplier: supplies hardware \added{- e.g. HP, DELL.}
\end{itemize}
An organisation can play one or several roles simultaneously (for example, a company can simultaneously play both the roles of a CSP and a NOP).

The 3GPP has also standardised the orchestration and management functions for the life-cycle management of network slices. Specifically, the 3GPP slicing management architecture boils down to three essential management NFs, called CSMF, NSMF, and NSSMF, as also illustrated by Figure~\ref{fig:ns_mngmt}: 
\begin{figure}[ht]
\centering
\includegraphics[clip,trim= 3.5cm 4cm 13cm 8cm,width=0.35\textwidth]{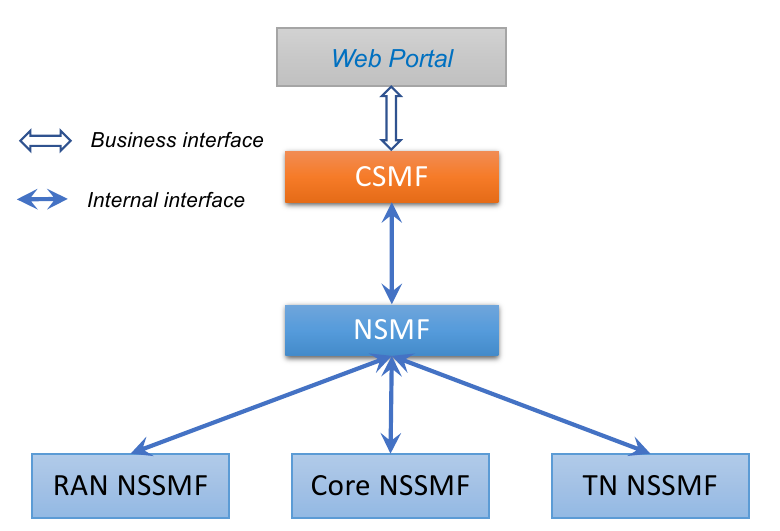}
\caption{3GPP network slice management architecture}
\label{fig:ns_mngmt}
\end{figure}
\renewcommand\labelitemi{$\bullet$}
\begin{itemize}[noitemsep,topsep=2pt]
    \item \textit{Communication Service Management Function} (CSMF). The CSMF is the user interface for slice management. It converts the SLAs of the CSs requested by the CSC into SLS and delegates the management of NSIs to NSMFs.
    \item \textit{Network Slice Management Function} (NSMF). The NSMF  manages NSIs and splits them into subnets for the RAN, transport and core domains. Then, the NSMF delegates the management of slice subnets to NSSMFs
    \item \textit{Network Slice Subnet Management Function} (NSSMF). The NSSMF applies the NSMF's life-cycle management commands (e.g., instantiate, scale, terminate, remove) within a particular subnet. 
\end{itemize}
Then, a 3GPP tenant can request the creation of a network slice from the NSTs made available by the CSMF, using a dedicated front end (such as the web portal depicted in Figure~\ref{fig:ns_mngmt}). It is also worth pointing out that the NSSMF is where most of the slice intelligence resides. It takes a command from the NSMF, such as "build a slice," and activates it by doing all the behind-the-scenes work of function selection, storage, configuration, and communication. Once each slice subnet is created, the NSMF is in charge of \textit{stitching} them together to build the end-to-end network slice.

\added{It is important to point out that SDN and NFV are the key technology enablers of network slicing in 3GPP networks. There is a plethora of open-source frameworks that offer software-defined implementations of the RAN and CN portions of the 4G/5G cellular architecture, which include Open Networking Foundation (ONF)'s projects}\footnote{\url{https://opennetworking.org}.} \added{such as Aether, SD-CORE, and SD-RAN, and solutions defined by the O-RAN Alliance}\footnote{\url{https://www.o-ran.org/specifications}.} \added{In addition several open-source projects have been established to implement the ETSI NFV Management
and Orchestration (MANO) framework~\cite{2017_JCSI_mano} to support the instantiation, control and configuration of network slices in different portions of the 5G infrastructure, such as ONAP, OSM and OpenBaton~\cite{2020_JSAN_mano}. The interested reader is also referred to~\cite{2020_COMNET_programmable_5G} for a comprehensive survey with extensive details on open virtualisation and management frameworks for 5G networks.} 
\subsection{ETSI Multi-access Edge Computing (MEC)}
\label{sec:mec}
\begin{figure}[ht]
\centering
\includegraphics[clip,trim= 0cm 3cm 10cm 4cm,width=0.45\textwidth]{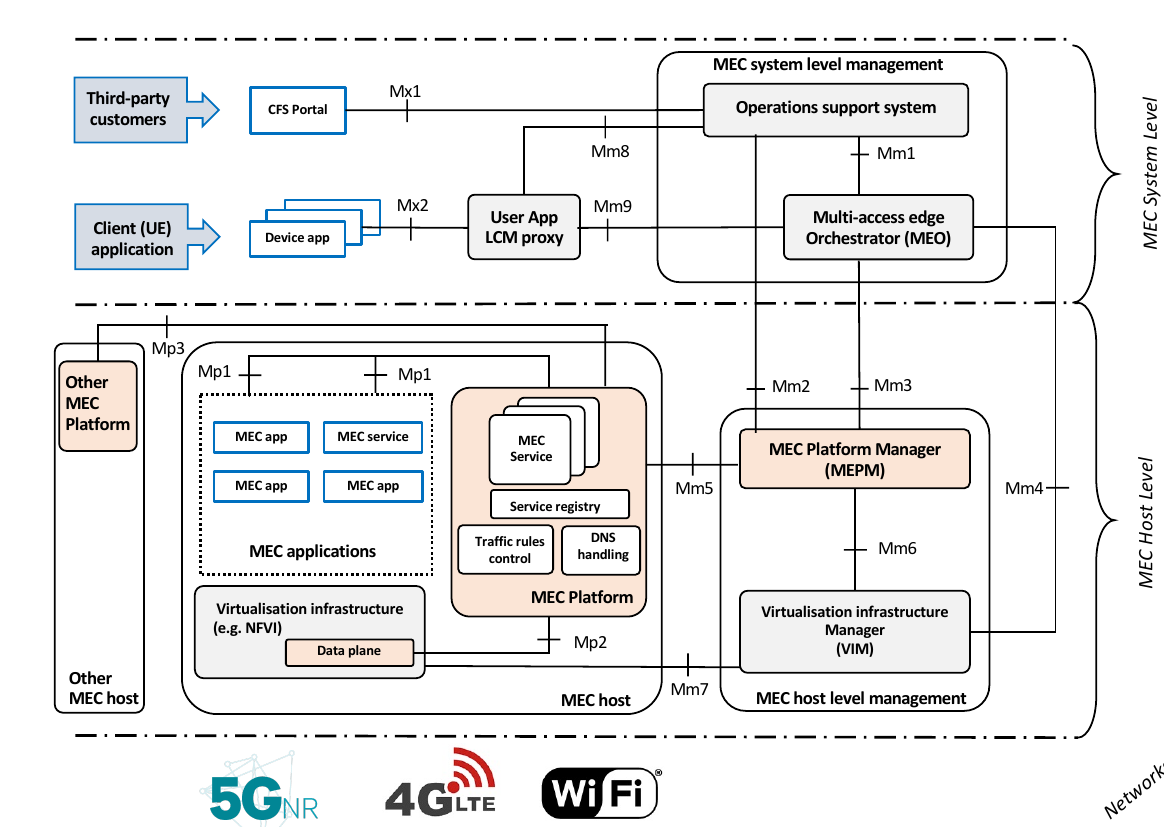}
\caption{MEC reference architecture (based on~\cite{MEC003}).}
\label{fig:mec_gen_arch}
\vspace{-0.2cm}
\end{figure}

\noindent
The ETSI organisation has introduced the Multi-access Edge Computing (MEC) since 2014 to provide a standard framework for the development of inter-operable applications over multi-vendor edge computing platforms. To this end, the MEC technology provides a new distributed software development model containing functional entities, services, and APIs, enabling applications to run on top of a generic virtualisation infrastructure located in or close to the network edge. For the sake of discussion,  Figure~\ref{fig:mec_gen_arch} shows the generic ETSI MEC reference architecture, which consists of three main blocks: $(i)$ the MEC Host, $(ii)$ the MEC Platform Manager (MEPM) and $(iii)$ the MEC Orchestrator (MEO). 

The MEC host is at the core of the MEC architecture as it contains: $(i)$ the generic virtualisation infrastructure (VI), which provides compute, storage, and network resources for the MEC applications; $(ii)$ the MEC applications running on top of the VI\footnote{Existing ETSI MEC specifications assume that MEC applications are deployed as VMs using an hypervisor-based virtualisation platform, but alternative virtualisation technologies and paradigms are under consideration~\cite{MEC027}.}; and $(iii)$ the MEC platform, an environment that hosts MEC services and offers to authorised MEC applications a reference point to discover and consume MEC services, as well as to announce and offer new MEC services. As discussed in detail in the following section, MEC services are an essential component of a MEC system, as they allow MEC applications to be \textit{network-aware}. In addition, the MEC platform is responsible for configuring a local DNS server and instructing the data plane of the VI on how to route traffic among applications, MEC services, DNS servers/proxies, and external networks. 

A management layer is associated with the MEC hosts of a MEC system, including a virtualisation infrastructure manager (VIM) and a MEC platform manager (MEPM). The former is responsible for allocating, maintaining, and releasing the virtual resources of the VI. The latter is responsible for managing the life-cycle of MEC applications and informing individual MEC platforms about application rules, traffic rules, and DNS configuration. Finally, the MEO is the core functionality of the MEC system-level management. Specifically, the MEO is responsible $(i)$ for selecting the MEC host(s) for application instantiation based on application requirements and constraints (e.g. latency), available resources, and available services; and $(ii)$ for maintaining an updated view of the MEC system. Furthermore, the MEO is interfaced with the Operations Support System (OSS) of the network operator, which receives the requests for instantiation or termination of applications from either applications running in the devices (e.g. UEs) or from third-party customers through the CFS portal. It is helpful to point out that the MEC standard provides a complete specification of only a limited set of necessary APIs, and associated data models and formats, but most of the management reference points are voluntarily left open by the standard to foster market differentiation. 

\added{It is important to point out that ETSI has established a dedicated Working Group, called DECODE, to promote the development of open-source MEC solutions that can offer all functional entities of the MEC architecture, or only a subset of them (for instance a MEC Platform, or an API implementation).}\footnote{A partial list of relevant MEC solutions is available at \url{https://mecwiki.etsi.org/index.php?title=MEC_Ecosystem}.}. \added{In addition, both academic projects and industrial initiatives are working towards the implementation of a variety of MEC-compliant edge computing solutions. Open-source implementations that are worthwhile to mention are: $i)$ LightEdge, a microservice-based implementation of the ETSI MEC architecture for 4G and 5G networks}\footnote{\url{https://lightedge.io/}.}\added{; $ii)$ LL-MEC, a platform built upon the OpenFlow and FlexRAN protocols that provides real-time radio network information to MEC applications}\footnote{\url{https://mosaic5g.io/ll-mec/}.}\added{; and Smart Edge Open, a software toolkit developed by Intel to build a variety of edge platforms, including MEC-complaint edge solutions embedded into 5G network}\footnote{\url{https://smart-edge-open.github.io/docs/product-overview/}.}\added{.}  
\subsubsection{MEC and network slicing}
\noindent
As pointed out above, the MEC is a distributed computing environment at the edge of the network, on which multiple applications can be served simultaneously while ensuring ultra-low latency and high bandwidth. To achieve this goal, applications have real-time access to network information through APIs exposed by MEC services. According to the MEC standard~\cite{MEC003}, each MEC system is mandated to offer services providing authorised applications with $(i)$ radio network information (such as network load and status), and $(ii)$ location information about UEs served by the radio node(s) associated with a MEC host. Furthermore, the \texttt{Bandwidth Manager} service, when available, permits both the allocation of bandwidth to certain traffic routed to and from MEC applications and the prioritisation of that traffic, also based on traffic rules required by applications. Based on the above, it is straightforward to realise that the fundamental design behind the MEC architecture is to enable a network-aware application design, namely, to allow MEC applications and MEC platforms to leverage network information to satisfy their requirements.

On the contrary, the 3GPP-based network slice concept envisions an architectural shift as it relies on an \textit{communication service-centric} network provisioning. For this reason, the ETSI MEC group has recently started discussing which new MEC functionalities and interfaces, as well as extensions to existing MEC components, are required to support network slicing, e.g. by including network slice ID into different MEC interfaces. The ETSI report~\cite{MEC024} has identified several use cases based on the different network slicing concepts that are advocated in different SDOs. 

\begin{figure}[htbp]
\centering
\includegraphics[clip,trim= 2cm 3.5cm 4cm 9cm,width=0.48\textwidth]{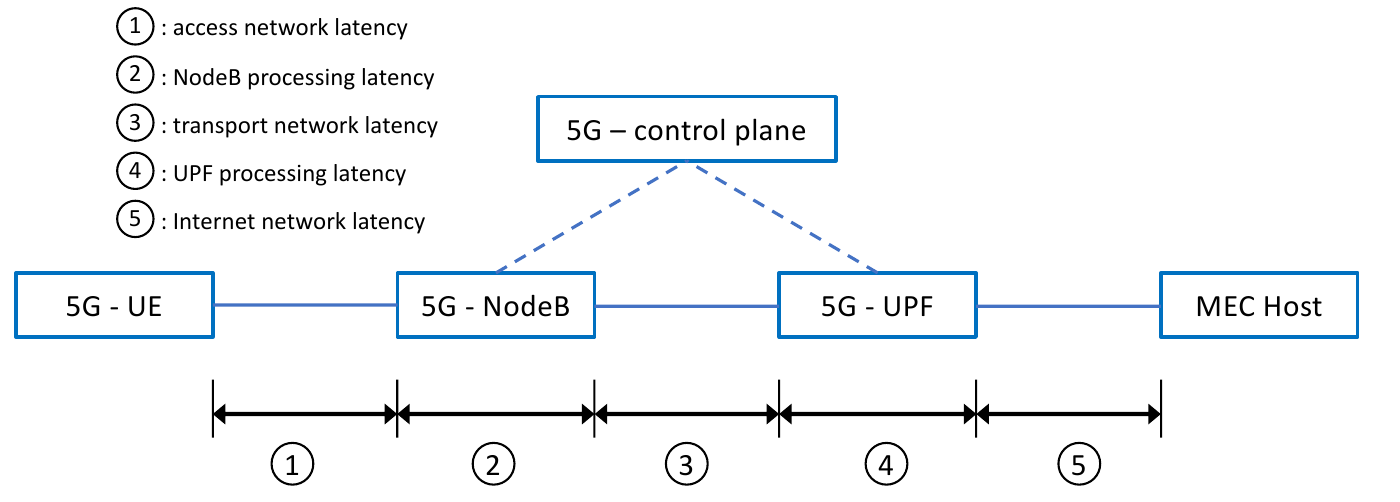}
\caption{End-to-end network latency of a NSI that include MEC according to~\cite{MEC024}.}
\label{fig:e2e_latency}
\vspace{-0.2cm}
\end{figure}
The most advanced proposal for supporting network slicing in MEC systems is the so-called \emph{MEC-in-NFV architecture} which permits to deploy MEC applications and MEC platforms as VNFs. To integrate MEC hosts in a NFV environment, ETSI standard proposes to substitute $(i)$ the MEPM with a MEPM-V entity that delegates the MEC-based VNF life-cycle management to one or more VNF managers (VNFM), and $(ii)$ the MEO with a MEC application orchestrator (MEAO) that delegates to the NFV orchestrator (NFVO) the resource orchestration for MEC VNFs. In~\cite{MEC024} different uses cases are also discussed in the context of the MEC-in-NFV architecture, for instance, to enable the sharing of a MEC host with several NSIs, or to allow MEC applications belonging to multiple tenants to be deployed in a single NSI. ETSI MEC also recognises the importance of the contribution of MEC applications to end-to-end latency. Thus,~\cite{MEC024} also presents a use case in which the MEC platform is included in a 3GPP NSI, and the end-to-end network delay budget takes into account MEC components and in particular, the delay from the data network to the MEC platform as shown in Figure~\ref{fig:e2e_latency}. In this case, the descriptor of the network services to be created in the NFV environment shall include the network latency requirement, which is distributed to the access network, core network, transport network and data network. This would require extending the data model of the NFV service descriptors to include the Application Descriptor, which contains fields to indicate the type of traffic to offload and the MEC service to consume. However, such a proposal has the drawback of requiring the 5G CSMF to translate application-service-related requirements into network-slice-related requirements. Indeed, SLA frameworks regarding application services are typically different from communication services (as per the 3GPP definition). For instance, data backup is often part of an application service SLA but is not part of a communication service SLA. Thus, radical changes in CSMF implementations would be needed to support such use cases.

A recent study~\cite{Cominardi2020} has proposed a harmonised view of the use cases defined in~\cite{MEC024}, to enable MEC frameworks to provide transparent end-to-end multi-slice and multi-tenant support by facilitating the sharing of MEC applications and MEC platforms among multiple NSIs. To this end, an inter-slice communication mechanism is proposed that automatically filters exchanged data between network slices installed on the same MEC facilities. The work in~ \cite{2020_MNET_MEC_subslice} focuses on 5G networks and presents extensions of the slice management architecture to enable the integration of MEC segment into the network slice as a slice subnet by introducing the MEC NSSMF component. Our work takes the architecture in~\cite{Cominardi2020} and~\cite{2020_MNET_MEC_subslice} as starting points and we envision a business model in which multiple tenants exist who are willing to pay to get isolated slices of network and edge computing resources.  However, we go one step further with respect to~\cite{Cominardi2020} and~\cite{2020_MNET_MEC_subslice} as we show how tenants can be provided with complete but isolated MEC stacks so that they can independently orchestrate their assigned resources to the served customers. Furthermore, in our management architecture a tenant is not necessarily an end customer, but, as explained in details in following section, it may also be a service provider that offers its services on top of a virtualised MEC instance (MEC as a Service provisioning model). 

\section{An E2E Network and Application Slicing Architecture using MEC}
\label{sec:app_slice}
\noindent
In this section, we first introduce the concepts of application services and application slices, and we instantiate these concepts in the context of MEC environments. Then, we discuss application slice from both operational and business perspectives by introducing operation and business roles, respectively, and relationships of the different entities involved in application slicing and service provisioning. We also discuss the different architectural use cases that correspond to the proposed model. Finally, we elaborate on an orchestration/management architecture for 3GPP networks integrating MEC systems that will enable such new use cases, and we propose an extended MEC reference architecture required to support the envisioned management architecture.   
\subsection{Preliminary concepts}
\label{sec:pre_concepts}
\noindent
As it could be sensed from both the introduction and the background sections, the 3GPP CSs cannot be straightforwardly used to model the services offered by an edge computing platform. For the sake of disambiguation, in this work, we call the services offered to the customers of an edge computing platform as \textit{Application Services} (ASs). The main reason is that CSs and ASs are designed to support different business purposes, therefore, they are regulated by different SLA frameworks. Specifically, \textit{CSs are dedicated to the transmission of data and network traffic}, while \textit{ASs are designed to support services which are generally not communication oriented in essence, but focus on data processing}, such as Augmented Reality (AR), Virtual Reality (VR), Ultra-High Definition (UHD) video, real-time control, etc. From the management perspective, it would not be sensible to manage ASs with the existing 3GPP CSMF. Some extensions or new management functions should be added to the overall management framework to translate application-related SLAs into application-related SLSs and network-related SLSs. 

In analogy to the 3GPP network slice concept, we define an \textit{Application Slice} (APS) as a \textit{a virtualised data collection system and a set of data processing resources that provides specific processing functions to support an Application Service and the associated SLA. The APS can possibly include Network Functions to ensure correct data communication among the previous processing functions}. A key aspect of the application slice concept is the support of \textit{isolation} not only of the resources that are used by each application slice but also of the management data and functions. Similarly to a network slice, the application slice is an end-to-end concept and can span multiple domains (e.g., device, edge platform, network). Furthermore, an application slice is not a monolithic entity. Following the well-known principles of service-oriented architectures, it is built as an \textit{atomic data-processing function} which has well defined functional behaviours and interfaces, called in this work \emph{Application Component Functions} (ACFs). We can now define an \textit{Application Slice Subnet} (APSS) as a set of at least one ACF and possible NFs supporting an application slice. It is essential to point out that application slices and network slices are not independent concepts, but they are strongly intertwined as application slices deployed at the edge of the network require dedicated network slices to deliver the massive amounts of data and network traffic that application services will typically produce. In the following sections, we will also discuss two potential distinct models to allow an application slice to use a network slice.  

The general concepts that we have introduced so far can be easily instantiated in the context of MEC, NFV and 5G systems. First of all, a MEC platform can be used to offer ASs, which can be deployed as conventional MEC applications and implemented using a collection of ACFs. Furthermore, the MEC system can be considered as one of the domains over which an application slice can span. Thus, the MEC management layer should be responsible for the orchestration and deployment of one of the application slice subnets that compose the E2E APS, hereafter denoted by \textit{MEC application slice subnet}. It is important to point out that a MEC application slice subnet includes one or more ACFs, but it can also include VNFs to support network functionalities. Furthermore, the tenant of a MEC system is not necessarily an end customer, but the MEC system can offer its services following a MEC-Platform-as-a-Service model.  

Finally, it is worth pointing out the differences between VNFs and ACFs which are designed for different purposes - network traffic processing for the former and data processing for the latter. Furthermore, in a multi-tenant MEC environment, like the one in~\cite{Cominardi2020} or~\cite{2020_MNET_MEC_subslice}, it is likely that the MEC applications implementing ACFs will not be orchestrated by the same entity that would orchestrate principal VNFs. Nevertheless, there are a lot of common points in terms of operation and management between a VNF and an ACF as both rely on the same set of virtualisation technologies for their deployment. Thus, we will reuse and extend both 3GPP management architecture for network slice and MEC-in-NFV architecture to support the proposed E2E application slice framework.
%
\subsection{High-level roles for application slice management}
\label{subsec:roles}
\begin{figure}[t]
\centering
\includegraphics[width=0.4\textwidth]{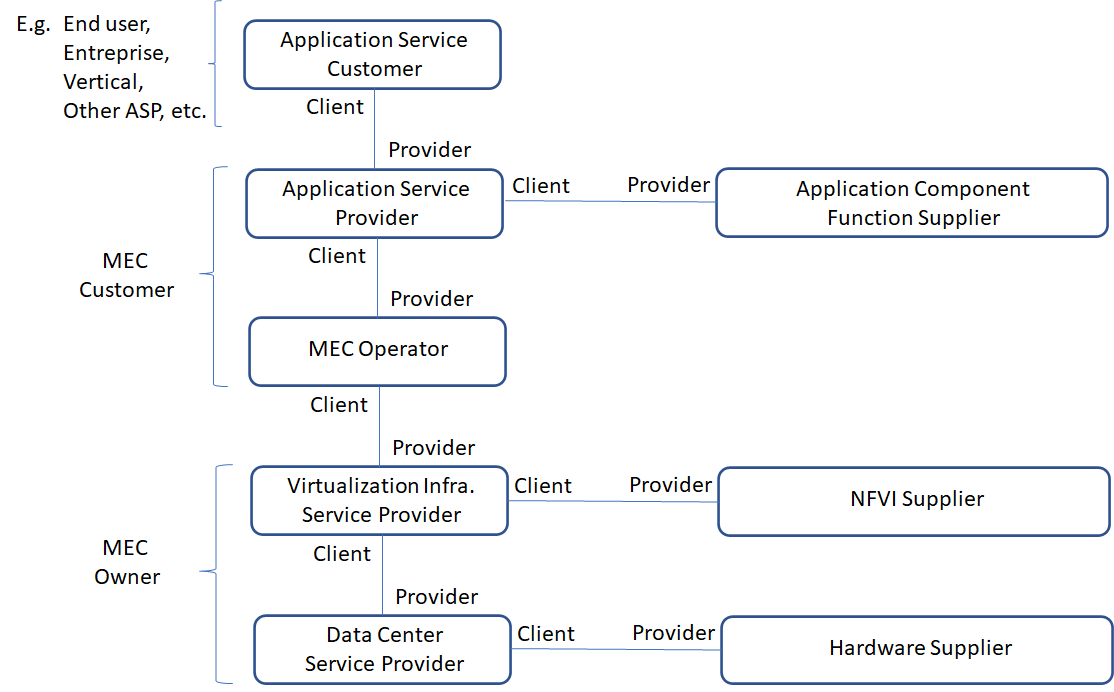}
\caption{High-level functional roles in the MEC application slice framework}
\label{fig:AS_roles}
\end{figure}

\noindent
From the APS management (and orchestration) perspective, we first need to define high-level roles in order to draw responsibilities boundaries similarly to what was proposed for 5G network slice management (see section~\ref{subsec:mgmt_architecture}). As discussed above, our focus is on Application Services that are offered by a MEC system. The Figure~\ref{fig:AS_roles} shows the different roles identified:
\renewcommand\labelitemi{$\bullet$}
\begin{itemize}[noitemsep,topsep=2pt]
    \item \textit{Application Service Consumer} (\textit{ASC}): uses Application Services.
    \item \textit{Application Service Provider} (\textit{ASP}): provides Application Services that are built and operated upon one or multiple application slice subnets, including MEC application slice subnets. Each of those application slice subnets is in turn built from a set of ACFs provided by the \emph{Application Component Function Supplier}.
    \item \textit{MEC Operator} (\textit{MOP}): operates and manages ACFs using a MEC Platform. We assume that the MEC Operator implements the MEC orchestrator and the ETSI MEC standardised interfaces as presented in~\cite{MEC003}. It designs, builds and operates its MEC platform to offer, together with the MEC orchestrator, MEC application slice subnets to ASPs from ACFs that the ASP has provided as an input.
    \item \textit{Application Component Function Supplier}: designs and provides ACFs to ASPs.
\end{itemize}
Interestingly, an ASC can become an ASP while adding new ACFs into consumed ASs and providing new ASs. For instance, let us assume that ASC $C_1$ uses from ASP $P_1$ an AS $S_1$ that provides face recognition capabilities. Then, ASC $C_1$ can integrate into $S_1$ other ACFs that permit to retrieve video stream from customers' cameras and ACFs that take control of customers' door, thus building a new AS $S_2$, which provides automatic door opening based on face recognition. ASC $C_1$ then becomes a new ASP $P_2$ selling the previous new AS to customers such as house/apartment rental platforms. Furthermore, we can easily make a parallel between Figure~\ref{fig:AS_roles} and Figure~\ref{2_figRoles} in Section~\ref{subsec:mgmt_architecture}. Roles responsible for low layers, namely Hardware Supplier, Data Center Service Provider, NFVI Supplier, and Virtualisation Infrastructure Service Provider, remain the same. However, the Application Component Function Supplier has replaced the VNF (or Equipment) Supplier; the MEC Operator has replaced the Network Operator, the Application Service Provider (ASP) the CSP, and the Application Service Customer (ASC) the CSC. 

It is helpful to point out that a business organisation can play one or multiple operational roles simultaneously. Therefore, without trying to be exhaustive in terms of business models, we focus in our work on two categories of organisations whose responsibility boundaries appear to make most of the sense as per our knowledge:
\renewcommand\labelitemi{$\bullet$}
\begin{itemize}[noitemsep,topsep=2pt]
    \item \textbf{\textit{MEC Owner}}: it plays both Hardware supplier and NFVI supplier roles. The fact that the MEC Owner also manages the virtualisation infrastructure and the edge servers allows him/her to dynamically split his/her infrastructure into logical partitions or network slices (i.e. greater degree of flexibility). It is noted that the MEC Owner could be the equivalent of what is called the \textit{MEC Host Operator} in \cite{MEC027}, as it offers virtualised MEC hosts and MEC platforms to its tenants. However, we prefer the `MEC Owner' terminology to avoid confusion with the `MEC Operator' role.
    \item \textbf{\textit{MEC Customer}}: plays both the role of the MEC Operator and ASP. It is helpful to point out that the ASP role offers a business interface (with the ASC). In contrast, the MEC Operator role offers the actual enforcement/implementation (i.e. SLS) of the business objectives (i.e. SLA) agreed across the ASP business interface. It is noted that the MEC Operator role alone cannot endorse a business organisation as it only offers Application Slice (including SLS) and not the Application Service (with related SLA). 
\end{itemize}
In conclusion, from a business perspective in this study, we advocate a multi-tenancy model in which ASCs are tenants of a MEC Customer, and MEC Customers are tenants of a MEC Owner. As explained in the following section, our proposed multi-tenant MEC system supports data isolation (through separated data planes) and resource orchestration separation (through separate resource orchestrators) between tenants.  
\subsection{Deployment models for an E2E application slice}
\label{subsec:AS_deployment}
\noindent
We distinguish two distinct deployment models for our proposed application slice concept: $(i)$ the \textit{overlay} model, and $(ii)$ the \textit{stitching} model. 

\begin{figure}[ht]
\centering
\includegraphics[width=0.5\textwidth]{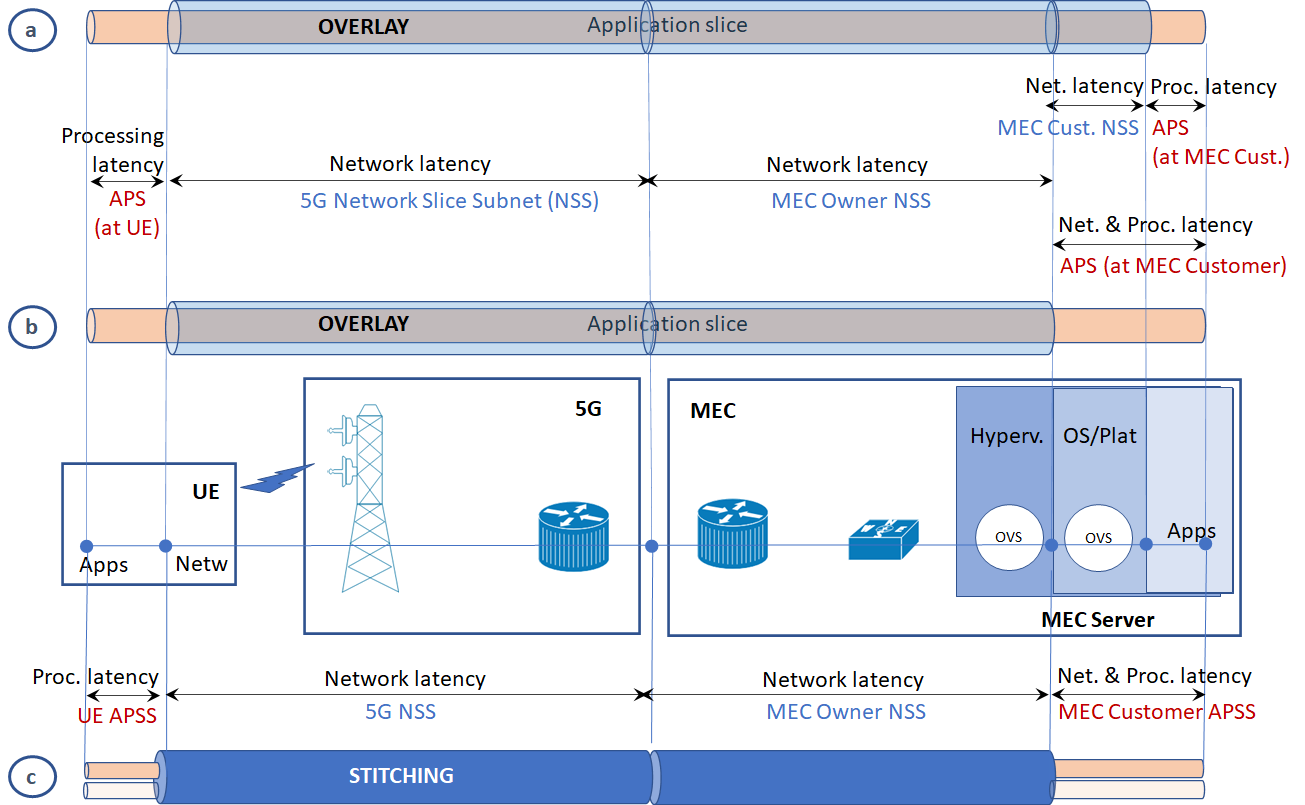}
\caption{\added{Interconnection models between application slice and network slice}}
\label{3_DeploymentModels}
\end{figure}
%
%
The first model assumes that the E2E APS is a consumer of a Communication Service offered by the underlying network slice (see Figure~\ref{3_DeploymentModels}\added{a}). In this case, the network slice is responsible for the entire communication latency, and we denote it as E2E \added{network} slice. In addition to the latter, the application slice caters for processing and possibly storage latency at both ends. It is important to point out that the E2E network slice encompasses not only the 5G network slice but also the network slice subnets within the MEC Owner and MEC Customer domains\footnote{Note that we use the term MEC Customer network slice subnet instead of MEC application slice subnet because, in this deployment model, the MEC Customer is only responsible for the management and orchestration of the data plane within its virtualised MEC environment. This is also justified by the need to align the architectural component boundaries with the liability boundaries of the deployment model.}. Indeed, as shown in Figure~\ref{3_DeploymentModels}a our slicing framework leverages recent advantages on virtualisation technologies that allow a virtualisation layer to be composed of multiple nested sub-layers, each using a different virtualisation paradigm. According to the functional role split illustrated in Section~\ref{subsec:roles}, a MEC Owner can use a hypervisor technology to operate its MEC hosts and to deploy multiple virtualised environments to MEC customers (e.g. allocating one or more VMs\footnote{In ETSI NFV terminology, a VM is also designated as `NFV VNF'.} using an NFVI). Each virtualised environment includes a full-fledged MEC system used by the MEC Customer to allocate further the resources assigned to its VMs to the multiple Application Services it deploys to its users by applying internal orchestration and management criteria. In the case of the MEC customer, a container-based virtualisation technology could be used as a top virtualisation sub-layer to manage application deployment into its allocated virtualised MEC system. It is worth noting that in \added{Figure~\ref{3_DeploymentModels}a} the E2E network slice includes the data plane of the MEC customer. However, an alternative overlay model is also possible in which the E2E network slice terminates at the network boundary between the MEC Owner and the MEC Customer as per \added{Figure~\ref{3_DeploymentModels}b}.

In the stitching deployment model \added{(see Figure~\ref{3_DeploymentModels}c)}, we assume that the MEC application slice subnet is a peer of the network slice subnets. Virtual appliances using a subnet border API, such as a virtual gateway or virtual load-balancer, can be used to interconnect MEC application slice subnets to the adjacent network slice subnet, similarly to what was proposed in~\cite{2020_MNET_MEC_subslice} and \cite{2021_TNSM_e22_slice_survey}. Such stitching could be a one-to-one interconnection as well as a multiple-to-one interconnection. The end-to-end application slice could be seen in this case as the composition of different application slice subnets (UE-operated or MEC-operated) together with network slice subnets. Latency-wise, the MEC application slice subnet is responsible, in this case, for a tiny part of the network latency budget in addition to the processing and storage latency induced by the MEC applications and their related ACFs.
%
%

We conclude this section by noting that the different deployment models will lead to different approaches to combine our proposed E2E application slice management/orchestration framework with the 3GPP management architecture, as explained in Section~\ref{subsec:AS_Mgmt}.

%

\subsection{Architecture for application slicing in a multi-tenant MEC system}
\label{subsec:new_MEC_arch}
\begin{figure*}[ht]
\centering
\includegraphics[clip,trim= 0cm 0cm 0cm 0cm,width=0.8\textwidth]{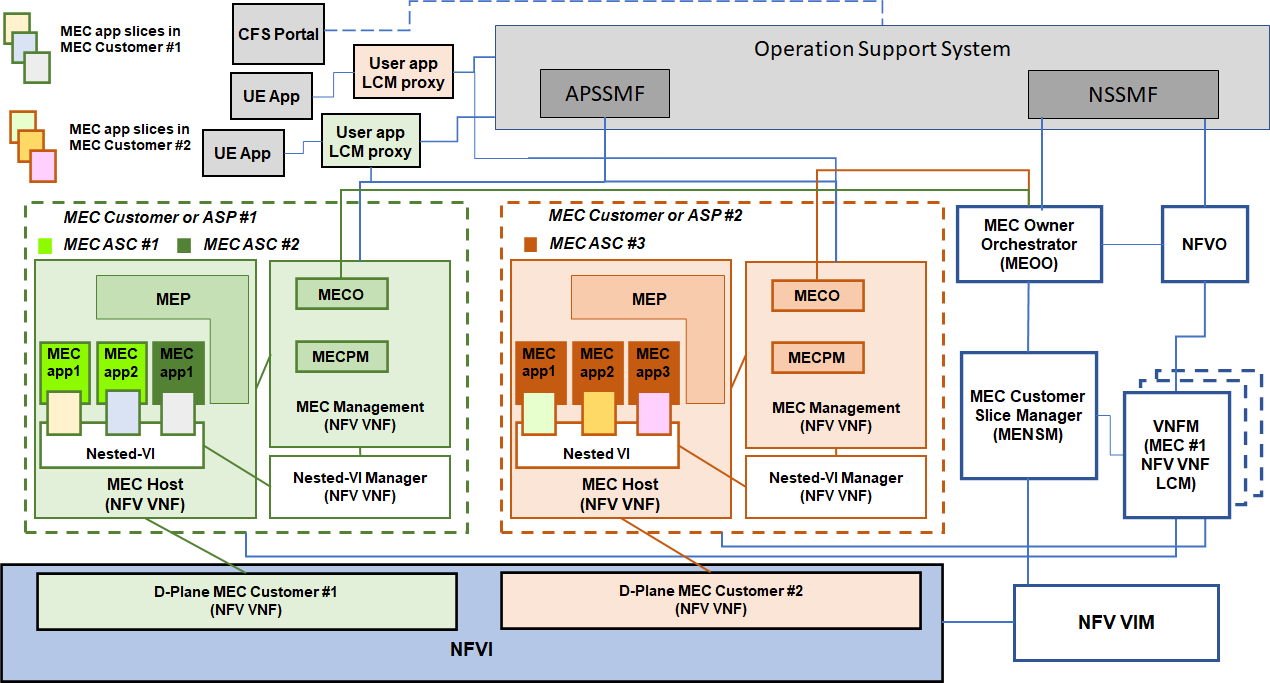}
\caption{Multi-tenant MEC architecture supporting network and application slicing}
\label{fig:ext_arch}
\end{figure*}

\noindent
In this section, we elaborate on the new MEC components and extensions to the current MEC management architecture that are needed to support E2E application slicing and multi-tenancy within multiple MEC customers. Figure~\ref{fig:ext_arch} shows an illustrative example of the proposed extended MEC reference architecture. The primary design rationale of our proposal is that the MEC system should be split into two responsibility domains following a \textit{two-layer hierarchical MEC architecture}, where the bottom layer is managed and orchestrated by the MEC Owner, and the top-layer is independently managed and orchestrated by MEC Customers. Such a hierarchical architecture allows a single MEC deployment to host multiple MEC Customers. Each of them has his own MEC network slice subnet (i.e. his dedicated data plane provided by the MEC Owner) with related management capability. In turn, each MEC Customer manages and orchestrates his own MEC application slices. 

Implementation-wise, the proposed two-layer MEC architecture is enabled by the nested virtualisation capability of the MEC infrastructure, as anticipated in Section~\ref{subsec:AS_deployment}. In the system illustrated in Figure~\ref{fig:ext_arch}, the MEC Owner does not deploy individual MEC entities as in the MEC-in-NFV reference architecture, but a collection of 'ETSI VNFs' (or VMs) to provide each MEC Customer with a complete MEC system. Differently from~\cite{Cominardi2020} we do not use 'ETSI VNFs' to deploy MEC applications and MEC platforms but to deploy a virtualised MEC environment encompassing virtualised MEC hosts and a virtualised MEC management system. Furthermore, in our MEC-in-NFV architecture variant, we introduce functional blocks that substitute the \added{MEAO} and \added{MEPM-V} of the original reference architecture. Specifically, we substitute the \added{MEAO} with the \textit{MEC Owner Orchestrator} (MEOO), which is in charge of implementing the policies to select the MEC infrastructures on which to deploy a MEC Owner network slice subnet. As explained in Section~\ref{subsec:AS_Mgmt}, the MEOO receives the commands to create, modify or delete a MEC Owner network slice subnet from a 3GPP management function called MEC NSSMF. Furthermore, the MEOO collaborates with the MEC Owner NFVO to provide a dedicated data plane to each MEC Customer. For the sake of example, we can assume that the MEC Owner offers to each MEC Customer a dedicated Kubernetes cluster, where each Kubernetes node is deployed as an 'ETSI VNF' (or VM) in the NFVI, which is connected to the 5G Core (5GC) via a dedicated data plane (MEC Customer network slice subnet). The second new functional block is the \textit{MEC Network Slice Manager} (MENSM), which delegates the life-cycle management of the 'ETSI VNFs' to a dedicated VNFM, while it is responsible for the management of the network slice subnet (data plane) parameters. For instance, it can reserve network bandwidth between MEC hosts for a given MEC Customer. More over, the MEC Network Slice Manager could behave like an 3GPP Application Function (AF) which interacts with the 5GC to synchronise data plane forwarding rules to realise local breakout traffics to/from MEC applications.

As previously discussed, each MEC Customer manages and orchestrates his own MEC application slices within the assigned virtualised MEC system. To this end, each MEC Customer implements a \textit{MEC Customer Orchestrator} (MECO), which receives the commands to create, modify or delete MEC application slices from a management function called MEC ASSMF (see Section~\ref{subsec:AS_Mgmt} below for more details). Furthermore, the MECO collaborates with the MEC Customer Platform Manager (MECPM) to manage the MEC application slice life-cycle and the MEC Platform instance (e.g. embodied as a containerised application). In order to stitch the application slice subnets to the adjacent network slice subnets, the MENSM creates dedicated VNFs (e.g. gateways) that it communicates to the MECO (or at least the gateway endpoints), similarly to~\cite{2021_TNSM_e22_slice_survey}. A collaboration between the MECO and the MEOO could also be needed in case of MEC application relocation to enforce new 5GC forwarding rules or to tear down old ones.

With regards to the interaction with the 5GC, there are two possible options:
\renewcommand\labelitemi{$\bullet$}
\begin{itemize}[noitemsep,topsep=2pt]
    \item The MEC Owner provides a network slice (i.e. a 'big pipe') to the MEC Customer, which directly manages via its MEC Platform (MEP) 5GC forwarding rules for each application slice (e.g. adds new DNS rules to the 5GC local DNS servers). This solution allows for better preserving privacy as the MEC Customer is the only entity that manages the data traffic produced by its own customers' UEs.
    \item The MEC Customer MEP collaborates (e.g. via the MECO and the MEOO) with the MEC Owner Network Slice Manager, which ultimately influences 5GC traffics. This solution allows for the MEC Customer to delegate the interaction with 5GC to the MEC Owner. The latter can aggregate requirements in order to optimise network resources (e.g. bandwidth). Thus, this solution allows for better network optimisation at the MEC Owner infrastructure but does not preserve privacy. Also, it may be less scalable as the number of UEs increases.
\end{itemize}
We conclude this section by noting that the standard MEC reference architecture~\cite{MEC003} entails a single MEC orchestrator controlling a single virtualisation infrastructure and managing the instantiation of all MEC applications. Our proposed MEC architecture variant implies a split of the MEC orchestrator responsibilities into a MEC Customer Orchestrator (MECO) and a MEC Owner Orchestrator (MEOO). While the former is responsible for MEC Platform, MEC applications, MEC application slices, and related external interfaces, the latter is responsible for the hardware, the NFVI, MEC NFVI slices (especially MEC network slices) and related external interfaces. 

\subsection{Application Slice Management Architecture}
\label{subsec:AS_Mgmt}
\begin{figure}[ht]
\centering
\includegraphics[clip,trim= 0cm 0cm 0cm 0cm,width=0.5\textwidth]{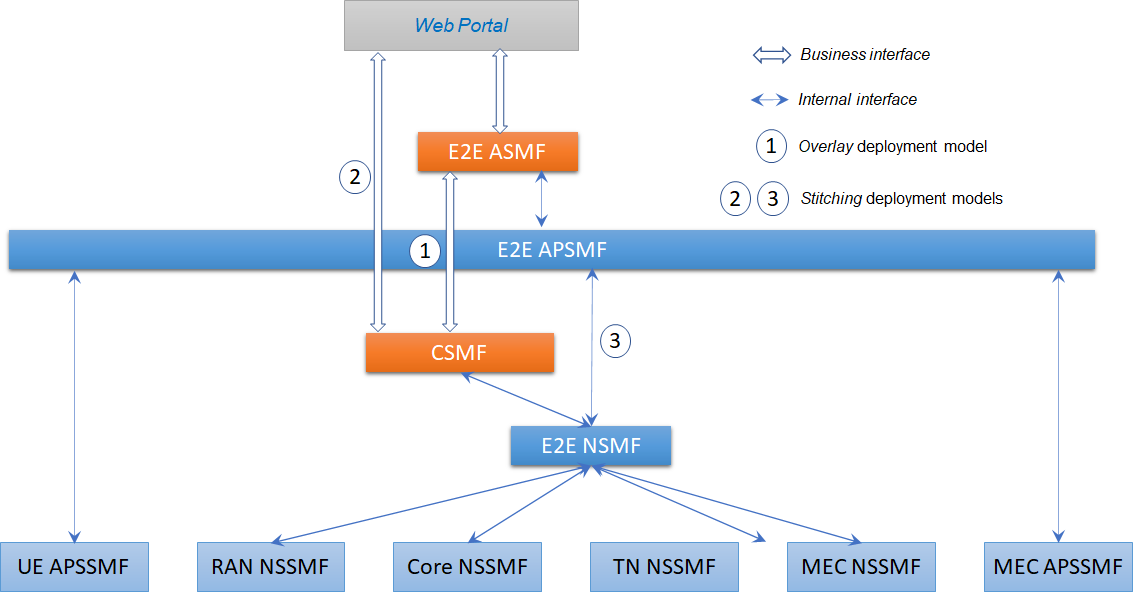}
\caption{3GPP-compatible joint network and application slice management architecture.}
\label{fig:aps_mngmt}
\end{figure}

\noindent
With the aforementioned new roles and architecture in mind, the 3GPP network slice management architecture could also be augmented to manage and orchestrate application slices as illustrated in Figure~\ref{fig:aps_mngmt}.
We assume that an ASC relies on a web portal to request an application service with a given SLA from a catalogue of offered ASs (see Section~\ref{sec:implementation} for more details on how to implement such service catalogue). The business interaction of the web portal can happen in two manners depending on the AS/APS deployment model that is used in the system (presented in Section~\ref{subsec:AS_deployment}). In both cases, the web portal communicates with a new management function, called \textit{Application Service Management Function} (ASMF), which is responsible for translating the SLA of the requested AS to the SLS of an APS and to trigger the creation of the APS instance by contacting a new management function called \textit{Application Slice Management Function} (APSMF). The APSMF splits the APS into multiple subnets, one for each domain over which the requested APS spans, including possibly the network and the AS's endpoints, namely the UE requesting the AS, and the edge system instantiating the AS. To this end, we introduce the new \textit{Application Slice Subnet Management Functions} (APSSMFs), which apply the APSS life-cycle management commands within the two potential domains that are relevant for an application slice subnets, namely the UE and the MEC.

In the overlay model (label 1 in Figure~\ref{fig:aps_mngmt}), the E2E ASMF is also responsible for translating the E2E AS SLA into E2E CS SLA and for requiring the adapted E2E CS from the CSMF. In the stitching deployment model, two alternative management flows are feasible. In the first case (label 2 in Figure~\ref{fig:aps_mngmt}), the end customer is an ``expert'' user and he is directly responsible for breaking (using the web portal) the E2E AS SLA into an E2E CS SLA and an SLA with a scope restricted to AS's endpoints. Then, the web portal is used for requiring the adapted E2E CS from the CSMF. In the other case (label 3 in Figure~\ref{fig:aps_mngmt}), the end customer is not an expert of the network domain and he does not need to perform the aforementioned E2E AS SLA splitting. On the contrary, the the APSMF is responsible to communicate directly with the NSMF to manage the network slice subnet associated with the APS. Finally, we remind that the management of network slice via the NSMF and the one of per-domain network slice subnets via NSSMFs are well-defined by 3GPP, and they do not need to be extended~\cite{3GPPTS28531,3GPPTS28541}. 

\begin{figure}[ht]
\centering
\includegraphics[clip,trim= 0cm 2.5cm 14cm 7.5cm,width=0.4\textwidth]{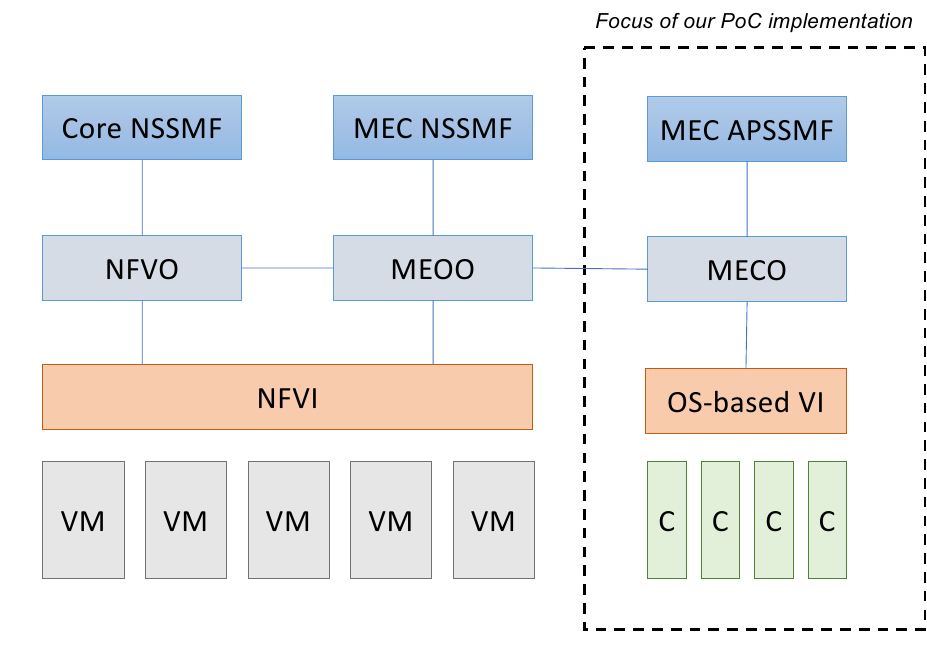}
\caption{Illustration of the orchestration entities that are involved in the MEC domain.}
\label{fig:ps_mngmt_focus}
\end{figure}
In the remaining of this paper, we will describe our experience in implementing the management architecture described above. As shown in Figure~\ref{fig:ps_mngmt_focus}, several orchestration entities are involved in the management of the various application slice subnets. Our focus will be on implementing the MEC Customer Orchestrator using a popular open-source container orchestration platform. Furthermore, we will detail the interfaces and data models needed to interact with the MEC APSSMF, allowing the deployment of isolated MEC application slice subnets composed of ACFs in the form of Docker containers.  


\section{Implementation Experience}
\label{sec:implementation}
\begin{figure*}[ht]
    \centering
    \includegraphics[clip,trim= 1cm 4.5cm 0cm 2.5cm,width=0.7\textwidth]{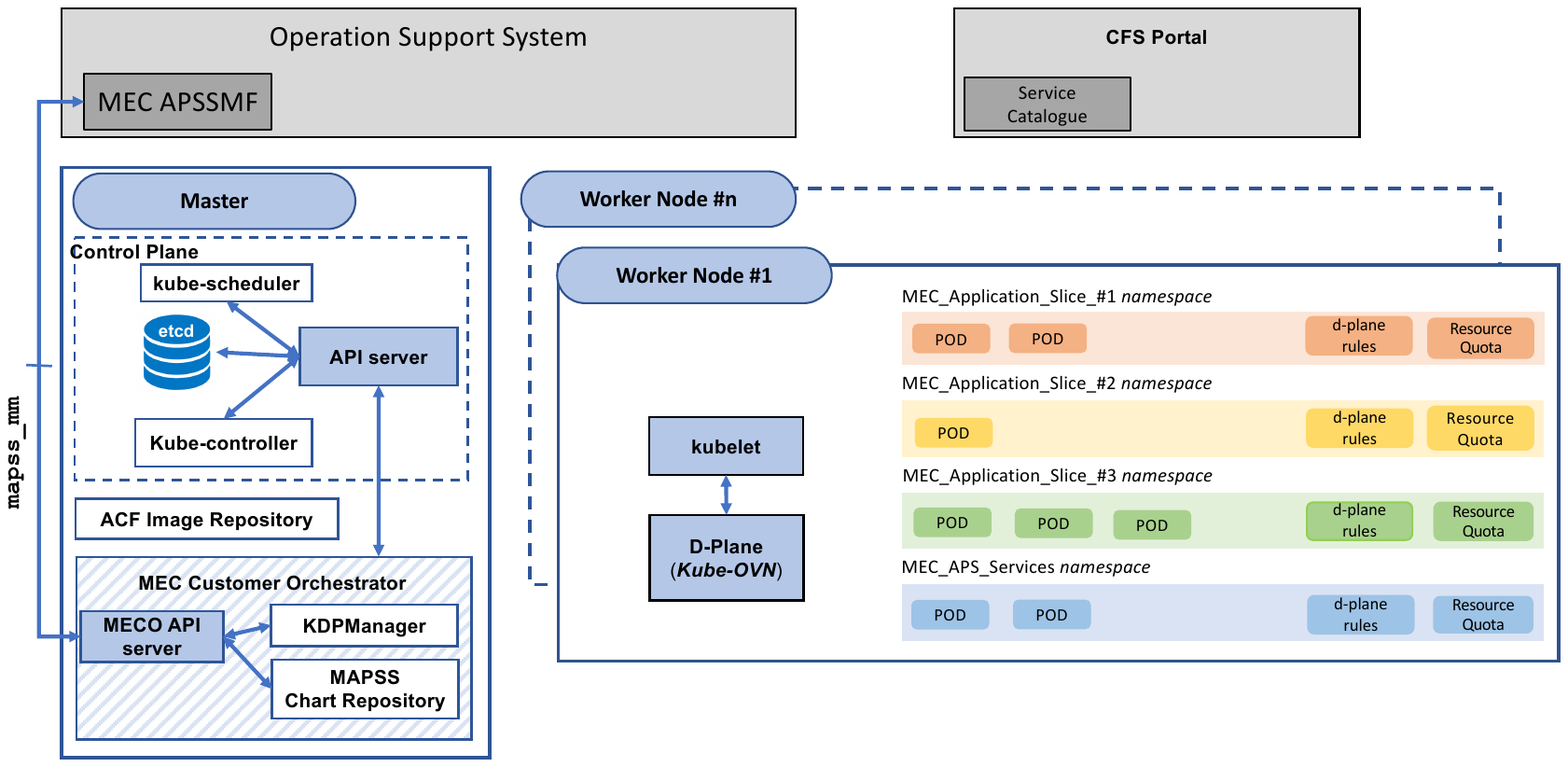}
   \caption{PoC architecture and internal components.}
    \label{fig:poc}
    \vspace{-0.2cm}
\end{figure*}

\noindent
In this section we describe our approach to implement a MEC Customer Orchestrator and to support MEC application slicing using Kubernetes and Helm technologies. Although our implementation is still at a work-in-progress state, our proof-of-concept prototype (hereafter simply PoC), shows the feasibility of our design approach and allowed us to collect a preliminary insight on the efficiency of application slicing using Kubernetes resource management capabilities. In the following, we first briefly overview Kubernetes and Helm features. Then, we detail the implementation of the new APIs and functional components  of our PoC,  namely the MECO and the ACF Image Repository. Finally, we describe our practical approach to support slicing of Kubernetes resources. For the sake of presentation clarity, Figure~\ref{fig:poc} overviews the architecture of the PoC and its internal components.
%
\subsection{PoC enabling technologies}
\label{sec:poc_tech}
\noindent
Kubernetes is an open-source container life-cycle manager and orchestrator, which is the de-facto standard for running container-based cloud native applications on a cluster of (physical or virtual) machines (called \textit{nodes}). It is out of the scope of this paper to describe the complete Kubernetes architecture and high-level abstractions, but we focus on the components that are most relevant for our PoC. 

A Kubernetes cluster is composed of $(i)$ a set of \textit{worker} nodes that run containerized applications, also called \textit{workloads}; and $(ii)$ (at least) a \textit{master} node that runs the services of the \textit{control plane}, and it is responsible to enforce the desired state of the cluster. Kubernetes provides several built-in workload resources to support various application behaviours (e.g., stateless tasks) and management functions (e.g., creating or deleting application replicas). Each workload must run into a \textit{Pod}, a Kubernetes object that represents a collection of containers running in the same execution environment, which share the same storage, networking, and lifecycle. Pods runs into worker nodes, which host an agent, called \texttt{kubelet}, that is responsible for managing worker's local containers and for synchronising the status with the master node. As better explained later, another main component of the worker node is the \texttt{kube-proxy}, which is responsible for implementing Kubernetes networking services and to enable communication to Pods from inside and outside the cluster. The master node is composed of different components including:$i)$ \textit{etcd}, a distributed key-value store that holds and manages all cluster critical data; $ii)$ \texttt{kube-apiserver}, a component providing a REST-based frontend to the control plane through which all other components interact; $iii)$ \texttt{kube-controller}, a component that monitors the shared state of the cluster using apiserver and runs the controller processes; and $iv)$ \texttt{kube-scheduler}, a component that assigns newly created Pods to nodes.  Note that Kubernetes also support the \textit{namespace} abstraction, namely virtual clusters that share the same IP Address and port space, facilitating the grouping and organisation of objects.    

Networking is a central part of the Kubernetes design and a fundamental capability for application slicing. Specifically, the Kubernetes network model demands certain network features, such as every Pod should have a unique, routable IP inside the cluster, and inter-pod communications should happen without using NATs, regardless of wherever the Pods reside or not on the same worker nodes (i.e. network segmentation is not allowed). Since IP addresses of Pods are ephemeral and change whenever the Pods are restarted or migrated, Kubernetes also defines \textit{Service} resources, namely REST objects that are used to group identical Pods together to provide a consistent means of accessing them, e.g by bounding them to a virtual IP address (called cluster IP) that never changes. It is important to point out that 
Kubernetes does not directly handle the networking aspects, but it rather allows the use of third-party networking plugins that adhere to the Container Network Interface (CNI) specification\footnote{\url{https://github.com/containernetworking/cni}} to manage the containers' data plane. Of particular relevance for our PoC, is Kube-OVN\footnote{\url{https://www.kube-ovn.io/}}, a CNI plugin that integrates network virtualisation into Kubernets  by leveraging the OVN (Open Virtual Network)\footnote{\url{https://www.ovn.org/en/}} technology. Kube-OVN supports advanced features, such as unique subnets per namespace, network policies, namespaced gateways, subnet isolation and dynamcic QoS. We extensively leverage some of those features to support application slicing in our PoC. 

Another technology we use as a basis for our PoC is Helm\footnote{\url{https://helm.sh/}}, an application packaging manager for Kubernetes. Specifically, Helm defines a data format, called \textit{Helm Chart}, to bundle a set of Kubernetes object definition files (i.e. files describing properties of Kubernetes objects) into a single package. This permits to manage the instantiation, upgrade and deletion of the corresponding Kubernetes objects as they were a single entity. The Helm charts are stored into a separate repository, and every time a new instance of the same chart is installed into Kubernetes, a new chart \emph{release} is created. Furthermore, Helm allows to augment Kubernetes object definition files with Helm template commands. By providing Helm a list of arguments for these template commands at chart instantiation time, it is possible to dynamically customise the chart before it is actually deployed.  Examples of these customisation span from overriding object default values with the one passed as command arguments (e.g. a port exposed by a container, the namespace name, etc..) to dynamically enabling disabling chart sections. As explained later, we extensively leveraged Helm features to implement different components of our PoC, such as the KDPManager and the MAPSS Chart Registry (see Figure~\ref{fig:poc}).

We conclude this section by noting that the ETSI MEC standard has recently started analysing how MEC features should be adjusted when deploying a MEC system using a container-based virtualisation technique~\cite{MEC027}. Furthermore, a few initial implementations exist of specific MEC components and interfaces using Kubernets as VIM, such as Akraino\footnote{\url{https://www.lfedge.org/projects/akraino/}} and LightEdge\footnote{\url{https://lightedge.io/}}. A recent work~\cite{2020_broadnets_mec_k8s} analyses how to use Kubernetes not only as a VIM, but also as the core of the MEPM, also leveraging Helm technology for the life-cycle management of MEC applications. 
%
\subsection{PoC design and development}
\label{sec:poc_design}
\noindent
One of the main objectives of our proof-of-concept implementation is to demonstrate how Kubernetes can natively support multiple, isolated instances of MEC application slice subnets (MAPSSs for brevity) as defined in Section~\ref{sec:pre_concepts}. The key building block of a MAPSS is the ACF. For the sake of simplicity, in our PoC we ignore VNFs and we assume that ACF Suppliers can provide ACFs in the form of Docker container images coupled with an ACF user manual or descriptor (called ACFD). Then, an ASP leverages the ACFDs to select the set of ACFs that are needed to build the AS requested by its end customers, as well as to derive the proper run-time configuration of the graph of ACFs composing the AS. Therefore, a key component of our PoC is the \textit{ACF registry}, where authorised ACFDs are published and stored. The ACF registry is implemented using the open-source Docker Registry 2.0 application\footnote{\url{https://hub.docker.com/_/registry}}, a storage and distribution system for named Docker images. A generic ACFD is structured into two separate sections. The first one details the RESTful APIs exposed by the ACF (OpenAPI\footnote{\url{https://www.openapis.org/}} is used to specify these APIs in a standard, language-agnostic format). The second section details how to properly configure the ACF parameters in order to control how the ACF will behave at run-time. The ACF parameter customisation is a crucial aspect to consider, especially in the context of ACF composition and automated orchestration. An example of ACF customisation is the selection of buffer sizes, which may influence the run-time behaviours and performance of ACFs, directly affecting the fulfilment of SLA requirements. For the sake of simplicity, in our PoC we use \textit{environmental variables} to pass configuration paramters to the running container images of ACF. Finally, it is important to point out that the ACF registry must be accessed also by the \texttt{kube-scheduler} agent to fetch the container images of ACFs to be deployed.

The other key component of our PoC is the MECO that we have implements as an additional component of the master node, using Go as programming language. Internally, the MECO component is composed of three different modules: $i)$ the \textit{MECO API Server} to enable communications with the MEC APSSMF; $ii)$ the \textit{MAPSS Chart Repository} to store the templates of Kubernetes deployment plans of MAPSS; and $iii)$ the \textit{KDPManager}, to manage the Kubernetes deployment plans of run-time instances of activated MAPSS. In the following, we elaborate on the purpose and operations of each module more in detail. 

\begin{figure}[th]
    \centering
    \includegraphics[clip,trim= 0cm 16cm 10cm 0cm,width=0.5\textwidth]{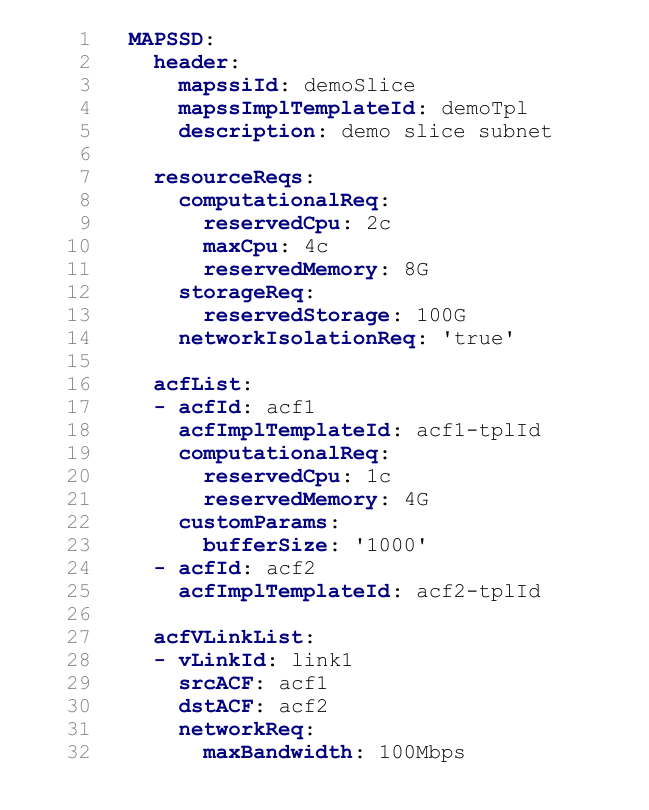}
   \caption{Illustrative example of a MEC application slice subnet descriptor (MAPSSD)}
    \label{fig:poc_massd_example}
\end{figure}

The main role of the MECO API Server is to act as a lifecycle management proxy for MAPSSs. Specifically, it receives commands for the instantiation, update and termination of MAPSS from the MEC APSSMF, and translates these commands into suitable Kubernetes actions. To this end, the MECO API Server exposes a RESTful management interface, called \texttt{mapss\_mm} API, defined using the OpenAPI description language (see Figure~\ref{fig:poc}). For the sake of the experimentation, the \texttt{mapss\_mm} API currently implements a \texttt{POST} method, which allows the MEC APSSMF to request the instantiation of a new MAPSS instance. The payload of this \texttt{POST} method carries a descriptor, called MAPSSD, which contains all the necessary information to allow the MECO to instantiate at run-time a specific MAPSS instance. The \texttt{mapss\_mm} API also implements a DELETE method, which allows to delete a running instance of a MAPSS. It is clear from this discussion, that the MAPSSD plays a key role in our PoC. We envision a MAPSSD organised into four different sections, as also illustrated in the example in Figure~\ref{fig:poc_massd_example}. The first parts includes the unique identifier of the MAPSS instance (\texttt{mappsiId}), a human readable description of the slice subnet features, and the identifier of an implementation template to be used for the deployment of the MAPSS instance (\texttt{mapssImplTemplateId}) -- see later this section for more details on how to use the \texttt{mapssImplTemplateId}. The second section includes a set of ``\textit{slice-subnet-wise}'' (computational, storage and networking) resource requirements. For example, the MAPSSD shown in Figure~\ref{fig:poc_massd_example} requires two dedicated CPU cores, 8 GB of memory and 100 GB of permanent storage, which will be shared among its constituent ACFs. The third part includes the list of ACFs that compose the MAPSS instance. Each ACF is associated to a unique identifier (\texttt{acfId}), and ``\textit{acf-wise}'' resource requirements can also be specified. For example, the acf1 in Figure \ref{fig:poc_massd_example} requires a dedicated CPU core out of the two dedicated to the whole slice subnet. Moreover, the field \texttt{customParams} can be leveraged to pass arguments (e.g. buffer size in figure) to configure specific ACF behaviours. Finally, the last section includes a list of virtual links among pairs of ACFs and their networking requirements (e.g. maximum usable bandwidth). It is worth noting that the data format of the MAPSSD is agnostic from the underlying virtualisation technology. In other words, the MEC APSSMF could leverage the same data model to interact with a MECO that relies on a different VIM than Kubernetes. Another advantage of the proposed MAPSSD is that it allows the MEC APSSMF to seamlessly integrate SLS requirements that address needs of different architectural levels (i.e. specific to the entire slice subnet, individual ACFs and virtual links between ACFs) in the same data object. 

Clearly, the translation from a VI-agnostic MAPSSD into a Kubernetes deployment plan (\textit{KDP} for short) of the MAPSS instance, namely a package of properly configured Kubernetes objects implementing the requested MAPSS instance, is a critical functionality of the MECO. Following the approach proposed in~\cite{NGMN028} for supporting cost-efficient customisation of network slices, we assume that the MECO hosts a pre-loaded set of MAPSS templates/blueprints that can be used to speed up the creation of a MAPSS instance. Specifically, we implemented each MAPSS KDP template as an Helm chart that includes a set of pre-configured Kubernetes objects. These objects define: $i)$ ACFs Docker containers to run (e.g. via Kubernetes Pods objects), $ii)$ ACFs behavioural parameters (e.g. via environmental variables in Kubernetes ConfigMaps objects), $iii)$ ACFs connection points (i.e. exposed ports), $iv)$ custom scheduling policies (e.g., number of replicas, failure behaviour, etc.), and any other Kubernetes feature that is necessary for the correct deployment of the MAPSS instance. Then, the \textit{mapssImplTemplateId} field of the MAPSSD is used to retrieve the correct MAPSS KDP template. It is important to point out that the MECO should be able to dynamically customise at run-time the MAPSS KDP template using information derived from the MAPSSD (e.g., container resource requirements). To this end we leverage built-in objects and control structures of Helm template that provide access to values passed into an Helm chart, and the ability to include conditions in the template's generation. In the current implementation, we limit such customisation to the selection of: $i)$ the name of the namespace to which objects will belong; $ii)$ the computational, storage and networking requirements for the namespace; $iii)$ the computational and storage requirements per ACF, and the networking requirements per ACF pairs. In our Poc the MAPSS KDP templates are stored in the MAPSS Chart Registry (see Figure~\ref{fig:poc}). According to the operational and management roles defined in Section~\ref{subsec:roles}, the MEC Customer plays the roles of both the MEC operator and the ASP. Thus, the MEC Customer has the necessary expertise not only to properly select, compose and configure ACFs to provide an AS, but also to select and properly configure the subset of Kubernetes objects that allows to implement the MAPSS instance of the designed AS. Finally, the KDPManager module is simply a wrapper of the Helm library, which allows the MECO to embed the Helm functionalities.   

\begin{figure*}[ht]
    \centering
    \includegraphics[clip,trim= 0cm 2cm 0cm 2cm,width=0.9\textwidth]{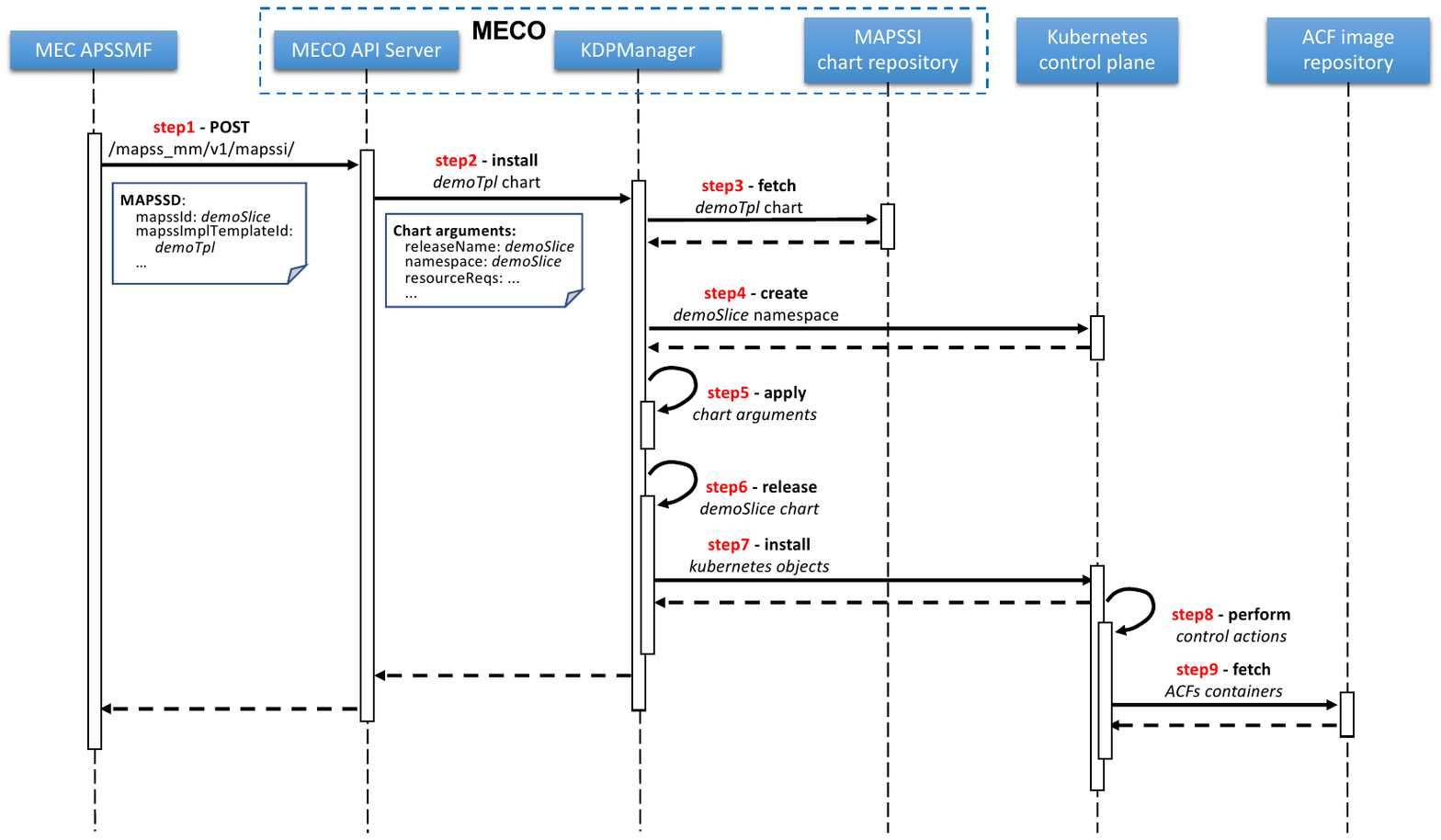}
   \caption{Sequence of operations to instantiate a new MAPSS in the PoC architecture.}
    \label{fig:poc_seq_diag}
\end{figure*}
We can now discuss the sequence of operations and request/response exchanges that are executed to deploy a new MAPPS instance in our Poc, which are also graphically illustrated in Figure~\ref{fig:poc_seq_diag}. First of all, the MEC APSSMF initiates the deployment process by sending a \texttt{POST} request to the MEC API Server over the \texttt{mapss\_mm} interface (see Figure~\ref{fig:poc}), which contains the MAPSSD of the requested MAPSS. In the figure, the requested MAPSS is identified as \textit{demoSlice}, while its Helm chart template is identified as \textit{demoTpl}. In \textit{step2}, the MECO API Server performs a preliminary analysis of the MAPSSD to discover the set of parameters that can be modified to customise the template. Furthermore, the MECO API Server retrieves from the \textit{mapssImplTemplateId} field of the MAPSSD the identified of the associated Helm chart template (\textit{dempoTpl} in this example). Finally, \textit{step2} is concluded with the API Server that instructs the KDPManager to deploy the \textit{demoTpl} Helm chart with the correct set of chart arguments. Subsequently, in \textit{step3}, the KDPManager fetches the \textit{demoTpl} Helm chart from the MAPSS Chart Repository, and it starts the deployment process. First, in \textit{step4}, the KDPManager contacts the \texttt{kube-apiserver} to create a new Kubernetes namespace with name \textit{demoSlice}. Then, the KDPManager applies (\textit{step5}) the customised arguments (e.g., number of cores to be assigned to an ACF container), and starts (\textit{step6}) chart release process, using \textit{dempoSlice} as release name. This process involves the generation of the proper set of Kubernetes objects definition files (i.e. the Kubernetes Deployment Plan). Once the KDP is complete, the KDPManager instructs the \texttt{kube-apiserver} to create the Kubernetes objects in the etcd database (\textit{step7}). Finally, in \textit{step8}, the \texttt{kube-controller} starts performing control actions according to the received Kubernetes objects. The latter includes contacting the ACF Image Repository to fetch ACFs container images for scheduling. For the sake of completeness, we remind that a termination of a run-time instance of a MAPSS is initiated by a \texttt{DELETE} request sent by the MEC APSSMF to the MECO via the \texttt{mapss\_mm} interface. This \texttt{DELETE} contains the \textit{mapssId}) of the MAPSS instance to delete. In this case, the MECO API Server requests the KDPManager to uninstall the chart release associated to \textit{mapssId}. Finally, the KDPManager removes all the Kubernetes objects of the release from Kubernetes cluster, and deletes the \textit{mapssId} namespace.

We complete the presentation of our PoC by explaining how application slice isolation is enforced using the Kubernetes resource control objects. First of all, we create a new Kubernetes namespace for each MAPSS instance, with a name equal to the \textit{mapssId}. All Kubernetes objects within a MAPSS instance are deployed using the same MAPSS namespace. We leverage a combination of Kubernetes \textit{ResourceQuota} objects, \textit{QoS classes} for Pods, and Kubernetes \textit{VolumeClaim} requests to limit the amount of computational and storage resources that could be consumed by both the whole namespace (to enforce requirements for invidual slice subnets) and individual Pods (to enforce requirements for individual ACFs). Network isolation between different instances of MAPSSs is implemented by exploiting Kube-OVN network policies, so that that the traffic arriving from Pods belonging to other namespaces is blocked (except for the system namespace). Finally, we implement per Pod ingress/egress rate limitation via Kube-OVN QoS policies. The implementation of more advanced QoS-aware network control policies (e.g. latency assurance, QoS tagging, etc.), and fine-grained network isolation policies (e.g. tunable network isolation degree with exception handling, etc) is planned as future work.

We conclude this section by  observing that an ASC could discover available ASes and their features by querying a catalogue that is exposed by a web portal (see Figure~\ref{fig:poc}), on which ASPs publish the descriptors of their ASes. We can foresee that AS descriptors include information such as $i)$ high level description of the offered AS; $ii)$ a pointer to the ASP offering the AS; $iii)$ a set of achievable SLAs (e.g. maximum resolution of a video processing service); and, possibly, $iv)$ billing information. The definition of a data model for AS descriptors is out of the scope of our present work. 
%
\subsection{Open implementation gaps}
\label{sec:poc_limitations}
\noindent
During the implementation of our PoC we also faced several difficulties due to the limitations of the technologies and standards we have used. In the following, we summarise the main technological gaps we have observed to highlight areas of future investigations.

\begin{itemize}[noitemsep,topsep=2pt]
    \item ETSI MEC specification has defined the methods and the data formats for the \texttt{Mm1} reference interface between the OSS and the MEO, which is used to trigger the instantiation and the termination of MEC applications in the MEC system. However, the \texttt{Mm1} implicitly consider a MEC application as a single application package. For instance, in the MEC-in-NFV architecture, the \texttt{Mm1} allows the MEAO to  deploy a single VNF onboarded as a VNF descriptor. In our use case, a MAPSS instance represents a set of ACFs, and it could be conveniently modelled as a graph. To some extent the \texttt{Mm1} interface should be expanded to resemble the capabilities of the \texttt{Os-Ma-nfvo} interface between the NSSMF and the NFVO~\cite{NFV-SOL005} , which allows the NSSMF to request a network service (i.e. a collection of VNFs) to the NFVO.
    \item Kubernetes ResourceQuota objects only permit to limit the amount of CPU and memory resources that Pods in a namespace could use. Pods can get assigned to a ``Guaranteed'' (highest priority) QoS class to receive reserved CPU and memory resources. VolumeClaim requests allow to reserve storage resources to a scheduled Pod. However, Kubernetes does not provide a straightforward mechanism to allocate resources at the namespace level but only at Pod level. This limitation complicates the implementation of resource over-provisioning strategies in dynamic slicing context (e.g. when a slice subnet is assigned more resources than needed to accommodate future demand changes).
    %
    \item Default scheduling mechanisms available in Kubernetes take into account only CPU and RAM usage rates when scheduling Pods, while network-related metrics (e.g. latency or bandwidth usage rates) are often ignored. However, a network-aware resource allocation and scheduling is crucial for our application slicing model, and initial proposals can be found in~\cite{2019_netsoft_netaware_kube} and~\cite{2020_noms_delayaware_kube}.
    \item Kube-OVN allows to limit the transmission rate on both ingress and egress traffic at the Pod level. This is obtained by using a QoS-aware queue and traffic policing at the vswitch port to which the Pod is connected. However, Kube-OVN does not support to set up rate limits on individual traffic flows, which is an useful feature if a Pod needs to communicate with several other Pods (e.g., for inter-slice communications). A possible workaround is to leverage Multus CNI\footnote{\url{https://github.com/k8snetworkplumbingwg/multus-cni}}, a CNI plugin for Kubernetes that enables attaching multiple network interfaces to pods, to allow a Pod to have a dedicated virtual interface (i.e. network port) for each destination Pod. Then, separated instances of Kube-OVN could be installed on each virtual interface to enforce different QoS policies at the port level. Furthermore, bandwidth reservation mechanisms similar to the ones proposed for SDN-based networks~\cite{2016_bwguar_openflow} should be included in Kube-OVN. 
    \item Service chaining allows to link together VNFs to compose service function chains (SFCs). The implementation of SFCs usually requires support from the network (e.g. via SDN) to route a packet from one VNF to the next in the chain. However, service chaining (which is a crucial feature for integrating VNFs with ACFs in a MAPSS is missing in Kubernetes. Recently, a few projects, such as OVN4NFV K8s Plugin\footnote{\url{https://github.com/opnfv/ovn4nfv-k8s-plugin}} and Service Meshes\footnote{\url{https://istio.io/}} have been initiated to provide support for service chaining in Kubernetes environments . 
    
    %
    \item The MAPSSD provides the blueprint for building an application slice subnet within a MEC environments. For the sake of our PoC, we have defined a custom data model for specifying a MAPPSD. On the other hand, standard modelling languages exist, such as TOSCA (Topology and Orchestration Specification for Cloud Applications) for describing the components of a cloud application and their relationships, which facilitate the interoperability, portability and orchestration in a multi-cloud environment~\cite{2018_MCC_TOSCA}. ETIS MANO already advocates the use of TOSCA to specify NFV descriptors~\cite{NFV-SOL001}. In principle, TOSCA could also be leveraged to specify the MAPSS descriptors. However, TOSCA is specifically designed to model classical cloud applications and it needs some adaptations to natively support also contenarised applications. Several approaches have been recently proposed to either $(i)$ extend the TOSCA normative types for support of container-based orchestration platforms (e.g. Cloudify\footnote{\url{https://cloudify.co/}}); or $(ii)$ to decouple the application modelling from the application provisioning by developing ad hoc software connectors between TOSCA workflow and cloud provider’s API (e.g. TORCH~\cite{2021_jgc_torch}). However, no standard specifications have been released yet.  
\end{itemize}

\section{PoC Experimental Validation}
\label{sec:evaluation}
\noindent
In this section, we first describe the evaluation setup and test bed infrastructure. Then, we introduce the application use case considered for the evaluation. Finally, we show the experimental results confirming the feasibility and efficacy of the proposed management architecture \added{applied to a cluster of VMs orchestrated via Kubernetes. The focus of our PoC implementation is to demonstrate that a MEC-specific orchestrator built upon advanced Kubernetes functionalities -- such as Kube-OVN to support OVN-based network virtualisation within Kubernetes -- can provide a bounded latency at the application level (i.e. at the Application Slice level) within the MEC domain if we can reserve sufficient resources for the Application Slice. Our experimental hypothesis is based of the fact that the E2E latency can always be apportioned into different domain latencies, namely the 5G RAN latency, the Transport latency, the 5G Core latency and the MEC latency. Then, each domain latency can be managed independently from domain-specific orchestrators, which coordinate to ensure the desired E2E latency.}
%
\subsection{Experimental Setup}
\label{sec:exp_setup}
\noindent
Our PoC is implemented using K3s\footnote{\url{https://k3s.io/}}, a lightweight Kubernetes distribution designed for resource-constrained edge servers. The software versions of all platforms used to develop our PoC and set up the Kubernetes cluster are listed in Table~\ref{tab:sw_version}. Our Kubernetes cluster is composed of a controller node (\textit{kctl}) and two worker nodes (\textit{kw1} and \textit{kw2}). Each Kubernetes node is a VMware virtual machine configured with Ubuntu as the guest OS, 4GB of RAM, 40GB of storage, but with heterogeneous computing capabilities. Specifically, \textit{kctl} node is provided with one virtual CPU, \textit{kw1} with three virtual CPUs and \textit{kw2} with four virtual CPUs. As explained in detail in Section~\ref{sec:use_case}, a video-analytics software is running only on \textit{kw2}. We deployed the MECO and the ACF Image Repository on \textit{kctl}. We populated the ACF Repository with the container images of the needed ACFs and the MAPSS Chart Repository with a set of Helm charts modelling the Kubernetes deployment templates of the MAPSS depicted in Figure~\ref{fig:poc_eval_slsub_diagram} (see next section for more details). K3s nodes are connected with virtual links having a bandwidth limit of 1 Gbps. Finally, the K3s cluster is running on a physical host equipped with an i7-7700 CPU$@$3.6GHz (4 cores, 8 threads), 16GB RAM, 500GB of storage. Therefore, each virtual CPU has a resource limit of 1000 millicores (mc). 
\begin{table}[t]
    \centering
    \small
    \begin{tabular}{|c|c|}
    \hline
        \textbf{Software} & \textbf{Version} \\
        \hline
         K3s & 1.21.5 \\ 
         Kube-OVN & 1.7 \\ 
         Go & 1.16.5 \\
         Docker & 20.10.6 \\
         Linux Kernel & 5.4.0.90 \\
         Operating System & Ubuntu 20.04.2 \\
         \hline
    \end{tabular}
    \caption{Software versions of the evaluation setup}
    \label{tab:sw_version}
    \vspace{-0.2cm}
\end{table}
%
%
\subsection{Use case}
\label{sec:use_case}
\noindent
%
%
\begin{figure*}[ht]
    \centering
    \includegraphics[clip,trim= 0cm 9.5cm 2cm 6cm,width=0.7\textwidth]{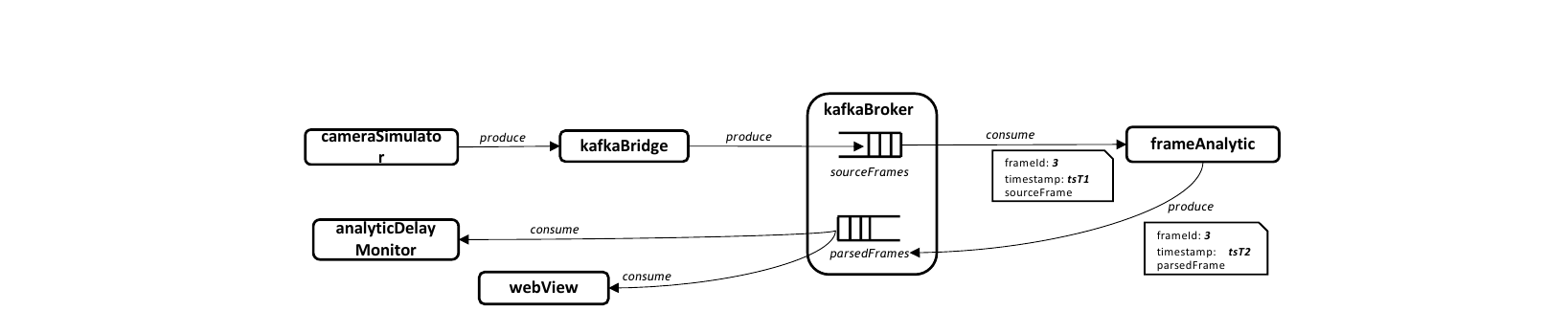}
   \caption{Graph representation of the ACFs and their interactions for the the deployed MAPSS.}
    \label{fig:poc_eval_slsub_diagram}
\end{figure*}
To validate the feasibility of our design approach we consider a use case from a typical Smart City scenario, where surveillance cameras are placed on particular points of interest or crowded areas and HD video streams as sent to an edge infrastructure where time-critical video analytic tasks (e.g. object recognition) are performed in a distributed manner. Specifically, we assume that a 5G-based Mobile Network Operator (MNO) has deployed a 5G infrastructure serving a smart city. In the 5G network premises, a MEC Owner also deployed a set of edge servers and manages a MEC virtualisation infrastructure. As a tenant of the MEC Owner, a MEC Customer leases a 5G network slice from the MNO and a sliced MEC environment from the MEC Owner to offer Video Analytic Application Services (VAASs) to its end customers. To this end, the MEC Customer has deployed a set of 5G-ready IP cameras across the city, and offers a variety of VAASs (e.g., traffic monitoring, pedestrian alerts, etc.) to different ASCs. (e.g., local municipality, firefighters, etc.). We assume that a VAAS request includes not only the category of the desired application service, but also a set of requirements (e.g. geographical scope, video accuracy, feedback timeliness, etc.). Then, the MEC Customer instantiates an E2E application slice, including a MEC application slice subnet, to satisfy the VASS request. 

Figure~\ref{fig:poc_eval_slsub_diagram} shows the service graph of the deployed MAPSS, namely the set of ACFs and their interactions. Without loss of generality, we assume that each ACF is implemented as a Docker container, and embedded into a single Pod. Therefore, ACF, Docker container and Kubernetes Pod concepts are used interchangeably in the following. In the deployed MAPSS, the \texttt{cameraSimulator} ACF is a data producer and transmits a video stream of images extracted from a public car dataset\footnote{\url{https://github.com/andrewssobral/vehicle_detection_haarcascades}}, \added{with a fixed frame rate of 25~fps}. We leverage Kafka for stream processing, and the \texttt{KafkaBroker} ACF embeds a standard Kafka broker. The \texttt{KafkaBridge} extracts individual frames from the video streams, compresses them, tags the compressed frames with a unique id (\textit{frameId}), appends a timestamp (\textit{tsT1}), and stores them with a custom \replaced{data}{dat6a} format in the \textit{sourceFrames} kafka topic of the \texttt{KafkaBroker}. \added{On average, every 40 ms a new frame is stored in the input queue of the \texttt{KafkaBroker}, which is configured so that stored messages that are not consumed within 2 seconds are removed.} The \texttt{frameAnalytic} ACF leverages the Python OpenCV library\footnote{\url{https://pypi.org/project/opencv-python/}} to perform vehicle detection on the stored frames. Specifically, \texttt{frameAnalytic} is subscribed to (read) the stream of events from the topic \textit{sourceFrames}. Then, vehicle detection is performed on each frame consumed by \texttt{frameAnalytic} and a new image is generated by drawing bounding boxes around every recognised car. Finally, processed frames are compressed, tagged with the same \textit{frameId} of the source image, timestamped (\textit{tsT2}), and stored in a new \textit{parsedFrames} kafka topic. The \texttt{webView} ACF is subscribed to the \textit{parsedFrames} topic and is responsible for displaying processed video frames on a web page. 
 
 To assess the performance of the deployed VASS, we use the \textit{application latency} of each frame, computed as the difference between the \textit{tsT2} and \textit{tsT1} values of frames with the same \textit{frameId}. The \texttt{analyticDelayMonitor} ACF is responsible for measuring processing delays. It is important to point out that we forced deterministic scheduling decisions in the \texttt{kube-scheduler} agent to ensure comparable delay measurements from different experiments. Specifically, Pods running the docker containers of the \texttt{frameAnalytic} ACF are always scheduled on \textit{kw2}, while all other Pods are assigned to \textit{kw1}. This also allows us to better investigate the impact of networking delays on the slice performance. 
%
\subsection{Experimental results}
\label{sec:results}
\noindent
In the following, we show results with the purpose of $(i)$ validating the ability of our management architecture to provide MAPSS performance isolation; $(ii)$ assessing the impact of resource constraints on the application latency\added{; and $iii)$ highlighting performance coupling between slices when resources are shared}. 
%
\subsubsection{MAPSS isolation}
\label{sec:test1}
\noindent
In the following experiments we deploy two identical instances of the MAPSS described in Figure~\ref{fig:poc_eval_slsub_diagram}, denoted as $slice_1$ and $slice_2$, respectively. All the Pods in $slice_1$ are assigned a Guaranteed QoS class\footnote{We remind that a Pod is assigned to the Guaranteed QoS class if for every container in the Pod, CPU and memory limits are equal to the CPU and memory requests. In Kubernetes, the request value specifies the min value the Pod will be guaranteed, while the limit value specifies the max value the Pod can consume.}, while all the Pods in $slice_2$ are assigned a BestEffort QoS class (i.e. they do not have any CPU/memory limits or requests). Table~\ref{tab:cpu_request} summarises the CPU requests of different ACFs. The CPU requests of ACFs deployed in $slice_1$ are decided after profiling the performance of each ACFs. Specifically, we run a preliminary experiment in which we deploy a single slice and we let the containers run without any resource limits. The values of CPU requests in Table~\ref{tab:cpu_request} represent the maximum CPU usage measured during this experiment.    
\begin{table}[ht]
    \centering
    \small
    \begin{tabular}{|l|l|c|}
    \hline
        ACF name & Worker & CPU request (mc) \\
        \hline \hline
        \texttt{cameraSimulator}  & \textit{kw1} & 100 \\
        \texttt{KafkaBridge}  & \textit{kw1} & 500 \\
        \texttt{KafkaBroker}  & \textit{kw1} & 500 \\
        \texttt{analyticDelayMonitor}  & \textit{kw1} & 100 \\
        \texttt{frameAnalytic}  & \textit{kw2} & 1800 \\
         \hline 
    \end{tabular}
    \caption{CPU requests (for Pods of $slice_1$) and scheduling plan for the different ACFs.}
    \label{tab:cpu_request}
    \vspace{-0.2cm}
\end{table}
To validate the ability of our solution to support MAPSS performance isolation (with a focus on CPU utilisation), we conduct an experiment with three distinct phases. In the first phase (from time 0 to time $T$), and third phase (from time $2T$ to time $3T$)) only $slice_1$ and $slice_2$ are active in the Kubernetes cluster. During the second phase (from time $T$ to time $2T$), an additional Pod is deployed on node \textit{kw2}. This ``interfering'' Pod runs within a single container a CPU stress tool that generates a CPU-intensive workload\footnote{\url{https://github.com/containerstack/docker-cpustress}}. Since, CPU resources for the Pods of $slice_1$ are guaranteed, this interfering Pod should only compete with the Pod of $slice_2$ running on node \textit{kw2}. To confirm our claim, Table~\ref{tab:avg_app_latency} shows the average application latencies measured during the three phases of the experiment, when $T=6$~minutes. 
\begin{table}[ht]
    \centering
    \small
    \begin{tabular}{|l|c|c|c|}
    \hline
         \multirow{2}{*}{Slice} &  \multicolumn{3}{|c|}{Application latency (msec)} \\\cline{2-4}
         & I phase & II phase & III phase\\
         \hline \hline
         $slice_1$ & 26.36  & 28.51 & 25.79 \\
         $slice_2$ & 27.00 & 85.66 & 25.78  \\
         \hline
    \end{tabular}
    \caption{Average application latencies measured during the various phases of the test.}
    \label{tab:avg_app_latency}
        \vspace{-0.2cm}
\end{table}
From the shown results, we can observe that $slice_1$ and $slice_2$ obtain similar application latencies in both phase I and phase III. This is sensible since $slice_1$ and $slice_2$ are assigned the same CPU resources by the \texttt{kube-scheduler}. Indeed, the Pod of $slice_1$ running on \textit{kw2} requests half of the available CPU capacity\footnote{We remind that \textit{kw2} has four virtual CPUs, which corresponds to 4000~mc. However, about 400~mc are used by the Kubernetes control-plane processes.} Therefore, the Pod of $slice_2$ running on \textit{kw2} is free to use the non-reserved resources without competing with other Pods. On the contrary during phase II, there is a competing Pod and the BestEffort QoS Class does not provide any guarantee on the assigned resources. Thus, we can observe that the application latency of $slice_2$ increases by a factor of 3.2, while the increase in application latency for $slice_1$ is negligible (only 8\%). \added{It is worth pointing out that no messages are lost in the message queues of the \texttt{KafkaBroker}. Thus, the average output throughput from the \texttt{FrameAnalytic} Pod is 25fps.}
\begin{figure}[ht]
    \centering
    \includegraphics[clip,angle=-90,trim= 2cm 0cm 2cm 0cm,width=0.50\textwidth]{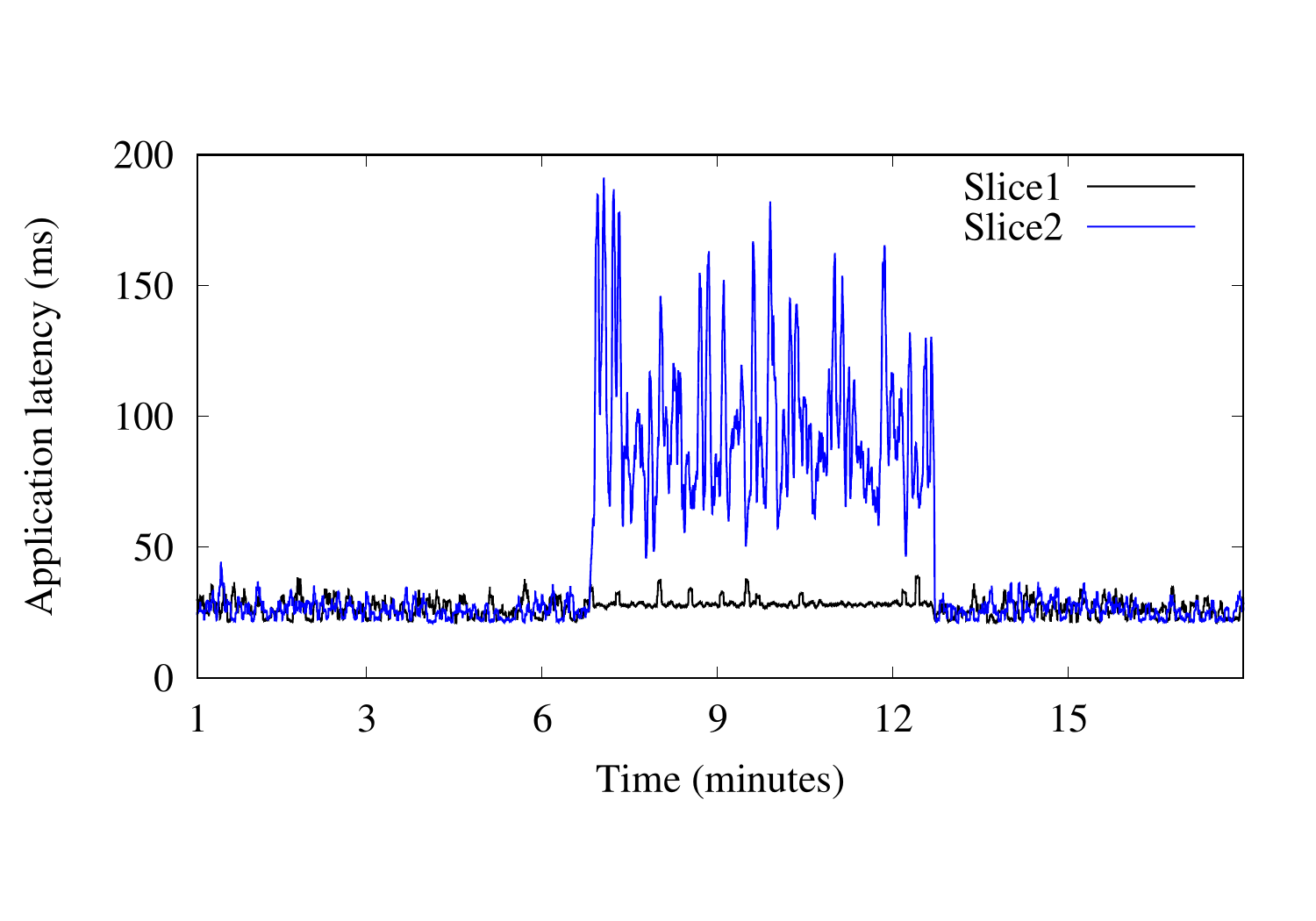}
   \caption{Temporal evolution of the applications latencies.}
    \label{fig:poc_slice_isolation_temp}
    \vspace{-0.2cm}
\end{figure}
We conclude this analysis by showing in Figure~\ref{fig:poc_slice_isolation_temp} the temporal evolution of the application latencies as observed in one of the experiments previously observed. The results not only confirm the findings of Table~\ref{tab:avg_app_latency}, but also show that the application latency values of $slice_2$ are affected by a notable variability. One main reason for these fluctuations of application latencies is that the frames in the video stream can generate different processing loads for the \texttt{frameAnalytic} Pod (e.g., depending on the number of vehicles to be detected). Therefore, if the \texttt{frameAnalytic} Pod is underprovisioned of CPU resources (as in the case of $Slice_2$), there can be spikes in the application latencies due to the \textit{sourceFrames} topic queue filling up. Another reason for the non-deterministic behaviours of application latencies is the way the standard CPU management policies work in Kubernetes. Specifically, time slices of a CPU over a fixed period are assigned to a process on the basis of CPU requests and limits. If the allocated number of slices in a period (100ms by default) are not sufficient to complete the application tasks, the process has to wait the following period to resume the processing, which causes a sudden increase of the application latency. Our PoC implementation approach, as described in Section~\ref{sec:poc_design}, is sufficient to limit the impact of these non-deterministic behaviours on $slice_1$. 
%
\subsubsection{Resource limitations}
\label{sec:test2}
\noindent
The purpose of the following tests is to assess the impact on application latency of reducing reserved resources with respect to the baseline setting used in the previous experiments. 
\begin{figure}[ht]
    \centering
    \includegraphics[clip,angle=-90,trim= 2cm 0cm 2cm 0cm,width=0.50\textwidth]{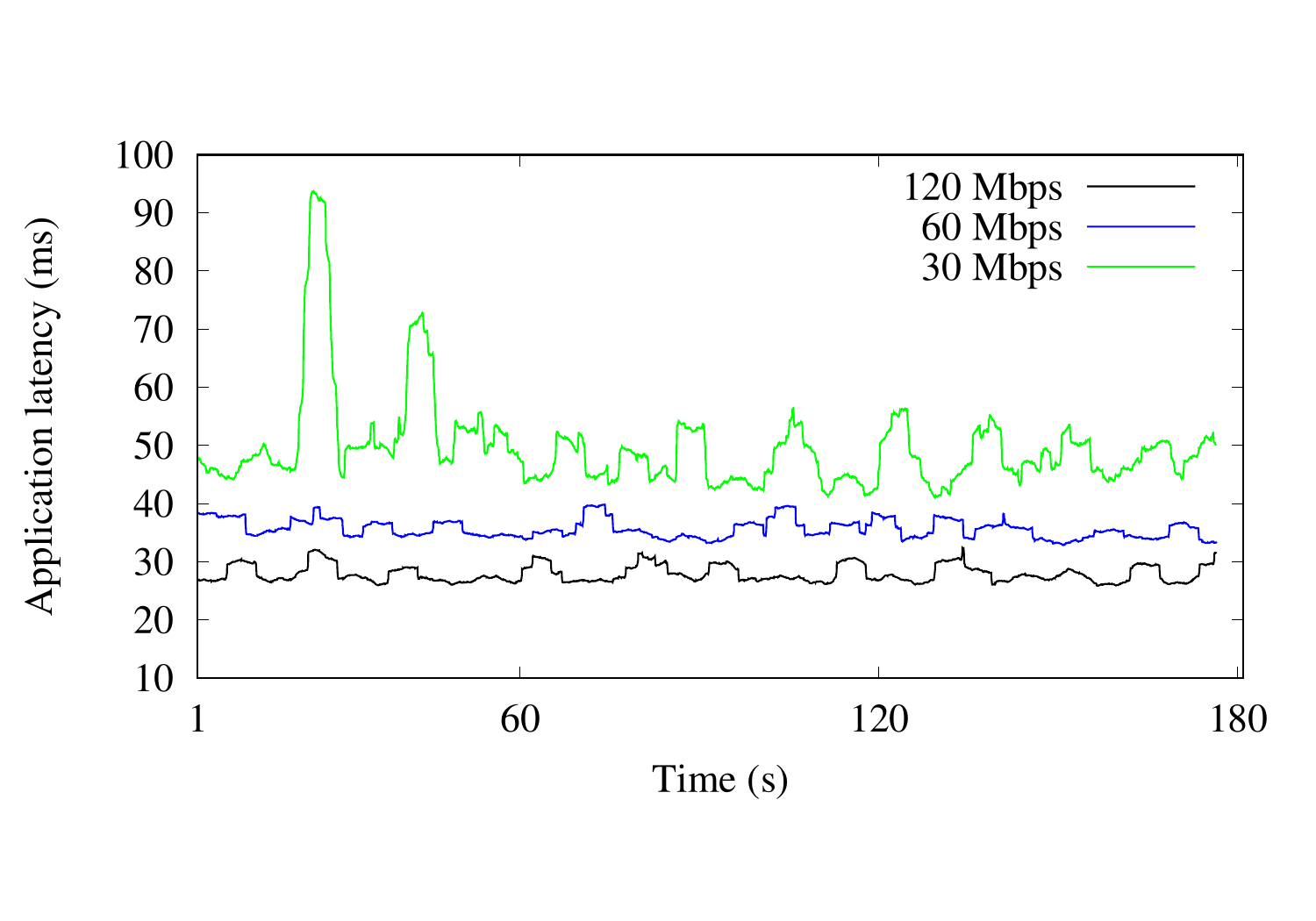}
   \caption{Impact on application latency of bandwidth constraints.}
    \label{fig:poc_res_bw}
    \vspace{-0.2cm}
\end{figure}
Figure~\ref{fig:poc_res_bw} shows how the application latency varies when reducing the bandwidth on the virtual link that connects the \texttt{KafkaBridge} ACF and the \texttt{frameAnalytic} ACF (in our experiemnatl setup this link is also the one connecting \textit{kw1} and \textit{kw2}). In the experiments we consider three bandwidth values $\{30,60,120\}$~Mbps. We can note that by halving the link capacity from 120~Mbps to 60~Mbps the application latency increases by 62\% on average, while decreasing the link capacity by a factor of four the application latency increases by 230\% on average. Furthermore, a higher latency variability can be observed when the link capacity is low, as it is more likely that the \textit{sourceFrames} topic queue gets unstable (i.e., the time between two consecutive frame requests of the \texttt{frameAnalytic} ACF is longer than the inter-arrival time of unprocessed frames in the \textit{sourceFrames} topic queue). 
\begin{figure}[ht]
    \centering
    \includegraphics[clip,angle=-90,trim= 2cm 0cm 2cm 0cm,width=0.50\textwidth]{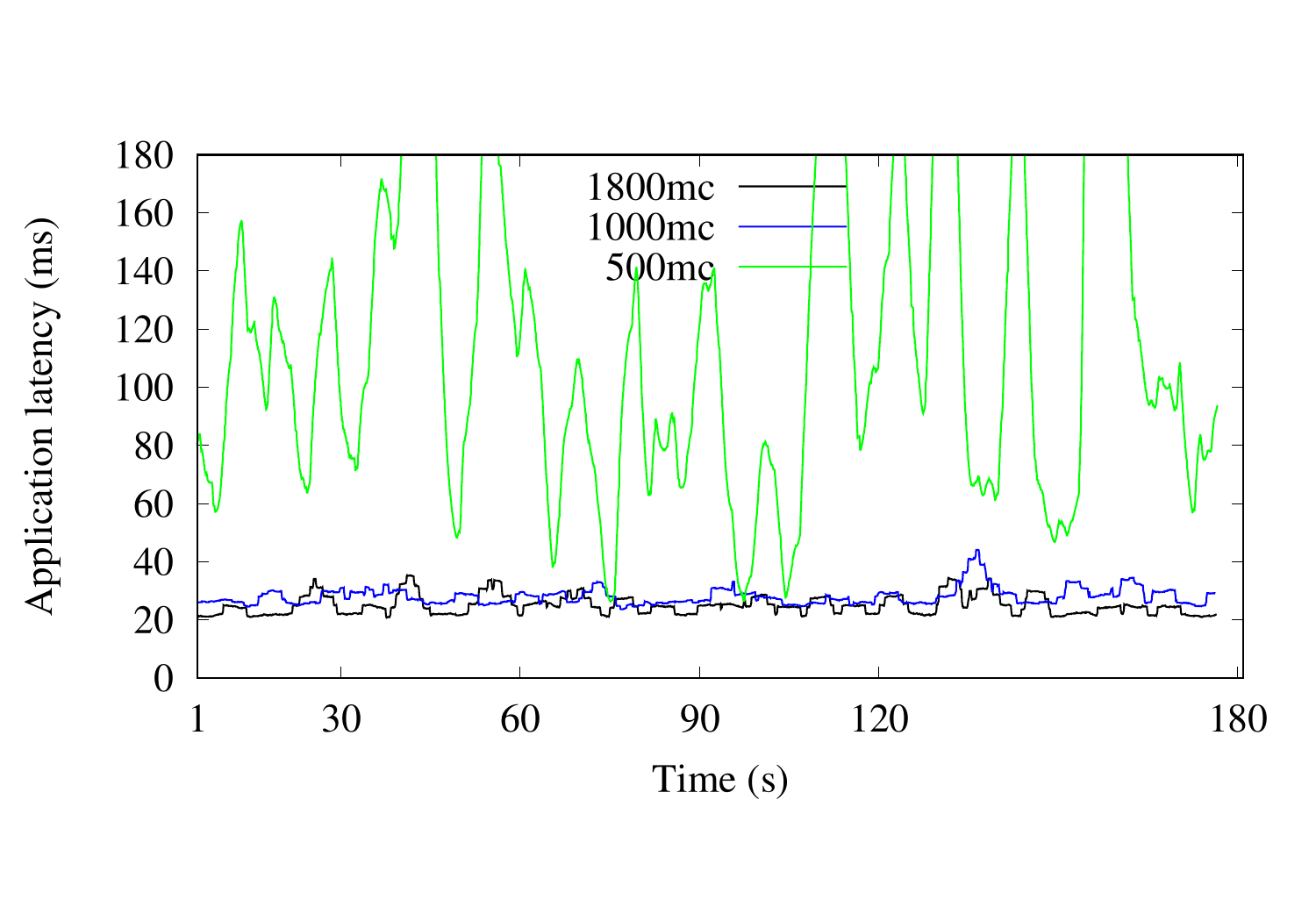}
   \caption{Impact on application latency of constraints on CPU speed.}
    \label{fig:poc_res_cpu}
    \vspace{-0.2cm}
\end{figure}
Figure~\ref{fig:poc_res_cpu} shows how the application latency varies when reducing the CPU quota of the \texttt{frameAnalytic} Pod. We can show that when we limit the CPU request of the \texttt{frameAnalytic} Pod to 1000~mc the average application latency only experiences a 5\% increase with respect to what observed with the base setting (namely a CPU request of 1800~mc). On the contrary, when the \texttt{frameAnalytic} is significantly underprovisioned (a CPU request of only 500~mc) not only the  application latency experiences a five-fold increase on average, but it also shows a very high variability. These results further confirm that a non-proper allocation of reserved resources may easily lead to non-deterministic behaviours due to a container scheduling process that is not aware of latency but only of CPU and memory requirements.
%
\subsubsection{\added{Shared edge resources}}
\label{sec:test3}
\noindent
\added{Typically, resource allocation in edge computing environments assumes that edge resources are dedicated to individual service requests. However, recent studies have started investigating more advanced use cases in which edge resources (e.g. the same set of data and code) can be shared between competing service requests. For instance, a joint service placement and data management solution for MEC-enabled IoT applications is proposed in~\cite{2021_jnca_edge} to take advantage of shared caches at the edge of the network. A similar problem is addressed in~\cite{2018_icds_edge_share} by assuming that service request can be satisfied by multiple replicas of the same service. Slicing algorithms that model the coupling between networking and MEC resources are also explored in~\cite{2020_mobihoc_si-edge}.}

\added{To evaluate the implications of using shared resources in our proposed slicing architecture we have built the following test. We deploy two identical slices, denoted as $slice_1$ and $slice_2$ that includes all the Pods described in Figure~\ref{fig:poc_eval_slsub_diagram} \textit{except} for the \texttt{frameAnalytic} Pod, which is deployed into a separate slice, denoted as $slice_3$. Then, the \texttt{frameAnalytic} Pod is \textit{shared} between $slice_1$ and $slice_2$. This is achieved by running two threads in \texttt{frameAnalytic} that are reading/writing from/to the \texttt{KafkaBroker} of $slice_1$ and $slice_2$, respectively. Another key difference with the tests described in Section~\ref{sec:test1} is that all the the Pods in $slice_1$, $slice_2$ and $slice_3$ are assigned a Guaranteed QoS class as we want a deterministic allocation of the maximum CPU time each Pod can use. In our experiments we consider two scenarios. In the first one, labelled \textit{with-sharing}, $C$ CPU units are guaranteed to the \texttt{frameAnalytic} Pod of $slice_3$. In the second one, labelled \textit{with-sharing}, the $slice_1$ and $slice_2$ deployment is the same as in Section~\ref{sec:test1}, i.e., two separate replicas of the \texttt{frameAnalytic} Pod are included into $slice_1$ and $slice_2$, but only $C/2$ CPU units are guaranteed to the two \texttt{frameAnalytic} Pods. In other words, in both scenarios the same total CPU processing power is allocated to the data analytic service. Table~\ref{tab:sharing} reports the applications latencies for different values of the CPU limit.} 
\begin{table}[ht]
    \centering
    \small
    \begin{tabular}{|c||c|c|c|c|}
    \hline
        \multirow{3}{1.8cm}{\added{CPU limit ($C$)}} & \multicolumn{4}{|c|}{\added{Application latencies (ms)}} \\
         \cline{2-5}
        &  \multicolumn{2}{|p{2.5cm}|}{\added{(\textit{non-shared serv.})}} & \multicolumn{2}{|p{2.5cm}|}{\added{(\textit{shared serv.})}} \\
        \cline{2-5}
        & \added{$slice_1$} & \added{$slice_2$} & \added{$slice_1$} & \added{$slice_2$} \\
        \hline
        \added{3600} & \added{25.95} & \added{26.01} & \added{19.39} & \added{19.65} \\
        \added{2000} & \added{26.22} & \added{26.20} & \added{21.93} & \added{21.63} \\
        \hline
    \end{tabular}
    \caption{\added{Average application latencies with shared and non-shared services.}}
    \label{tab:sharing}
    \vspace{-0.2cm}
\end{table}
\added{As expected, the higher the allocated CPU resources and the lower the application latencies. Furthermore, both slices achieve identical performance as their Pods are assigned a QoS class of Guaranteed. However, when the data analytic Pod is shared between the two slices, application latencies decrease up to 20\%. One main reason is that frames in the two video streams generate variable processing loads. Therefore, sharing the CPU resources reduces the impact of the processing time variability on the maximum buffer length (waiting time) of the \texttt{KafkaBroker}.}
%
%
%
\begin{figure}[ht]
    \centering
    \includegraphics[clip,angle=-90,trim= 2cm 0cm 2cm 0cm,width=0.50\textwidth]{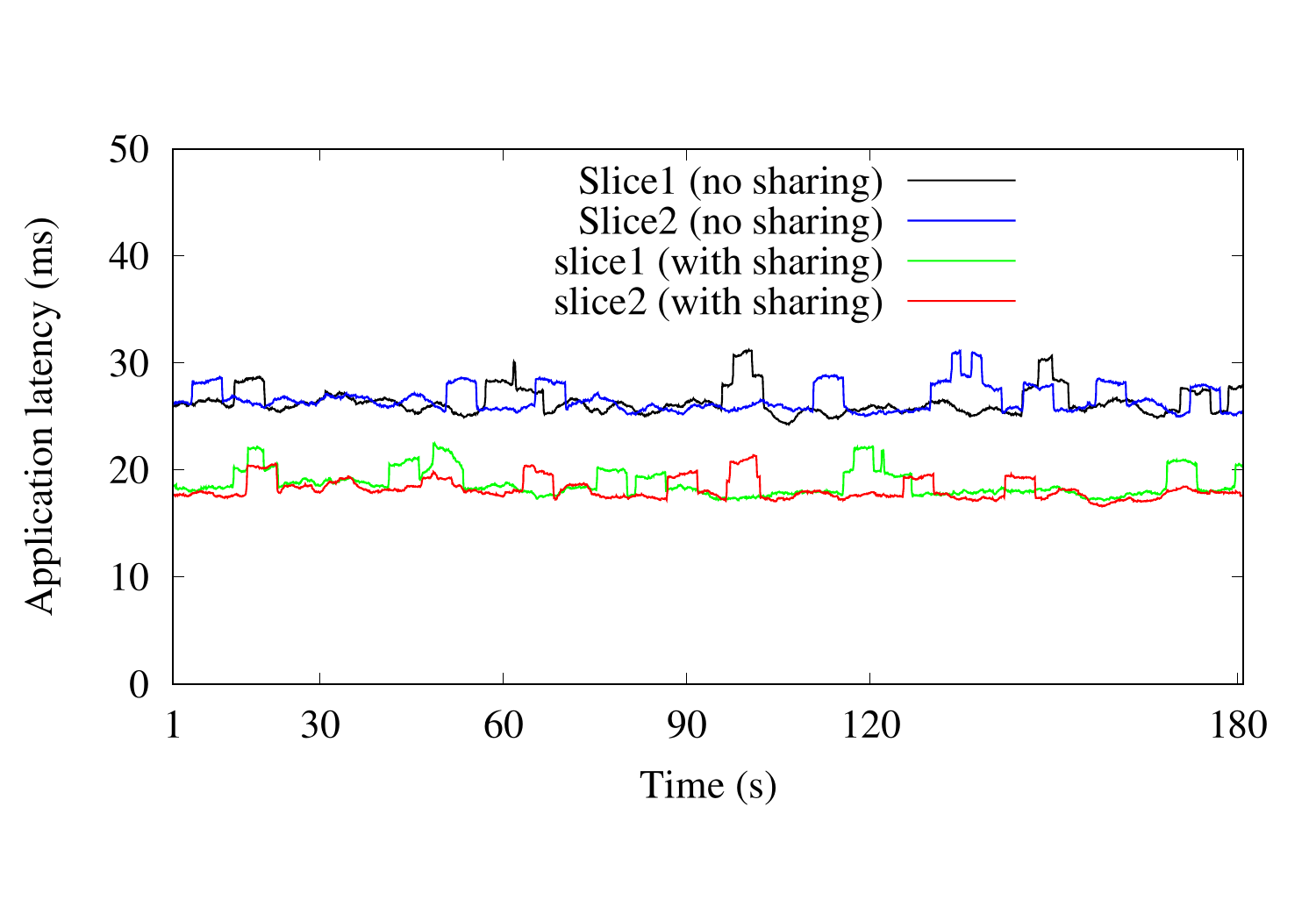}
   \caption{\added{Temporal evolution of applications slices with and without shared Pods (CPU limit $=$ 3600).}}
    \label{fig:sharing_1800}
    \vspace{-0.2cm}
\end{figure}
\added{Finally, Figure~\ref{fig:sharing_1800} shows the temporal evolution of the application latencies in the same experiment of Table~\ref{tab:sharing}. The results confirm the variability of the application latencies, but sharing the data analytic Pod ensure better performance than splitting the computing resource between two copies of the same Pod.}

\section{Conclusions\added{ and Future Work}}
\label{sec:conclusions}
\noindent
This article addresses the issue of integrating the 3GPP network slicing framework into the MEC infrastructure to facilitate the end-to-end management and orchestration of latency-sensitive resources and time-critical applications. To this end, we have proposed a new end-to-end application-centric slicing framework, which is built upon the new concepts of Application Service (AS), Application Slice (APS), and Application Component Function (ACF). Subsequently, we have designed a comprehensive multi-tenant MEC architecture for application and network slicing that considers different operational and business roles. We have also proposed a 3GPP-like integrated network and application slice management architecture compatible with a multi-tenant management architecture. The latter provides distinct management and orchestration responsibilities within each slice segment. Finally, we have proposed a way to implement a MEC Customer orchestrator within such a MEC slicing architecture, assessed the ability of our implementation approach to support performance isolation between applications, and discussed open implementation gaps. 

This work is the starting point for our future research on sliced, multi-tenant MEC infrastructures. Implementation-wise, there are still several architecture components to design with different possible design choices to benchmark. Architecture-wise, a solution to support MEC application relocation between different MEC Customers (i.e. between different tenants) is necessary. This solution would require a stronger interaction between our proposed MEC orchestration architecture and the 5G Core (5GC) network functions to synchronise traffic forwarding rules between multiple administrative domains (including MEC and 5GC). \added{Furthermore, several research problems are associated to the efficient sharing of resources (e.g., code, data, processing) between different slices. For instance, how to ensure not only traffic isolation (e.g., in terms of bandwidth), but also security isolation (e.g. data cannot be tampered between different slices) when sharing edge resources. Similarly, we plan to investigate techniques to efficiently aggregate and disaggregate shared resources when scaling up or down the different slices which use common shared resources.}
\vspace{-0.2cm}
%
%

\appendix


 \bibliographystyle{elsarticle-num} 
 \bibliography{cas-refs}

\begin{thebibliography}{10}
\expandafter\ifx\csname url\endcsname\relax
  \def\url#1{\texttt{#1}}\fi
\expandafter\ifx\csname urlprefix\endcsname\relax\def\urlprefix{URL }\fi
\expandafter\ifx\csname href\endcsname\relax
  \def\href#1#2{#2} \def\path#1{#1}\fi

\bibitem{2020_comst_5G_latency}
I.~Parvez, A.~Rahmati, I.~Guvenc, A.~I. Sarwat, H.~Dai, {A Survey on Low
  Latency Towards 5G: RAN, Core Network and Caching Solutions}, {IEEE
  Communications Surveys \& Tutorials} 20~(4) (2018) 3098--3130.

\bibitem{2021_access_urllc}
R.~Ali, Y.~B. Zikria, A.~K. Bashir, S.~Garg, H.~S. Kim, {URLLC for 5G and
  Beyond: Requirements, Enabling Incumbent Technologies and Network
  Intelligence}, {IEEE Access} 9 (2021) 67064--67095.

\bibitem{MEC003}
{ETSI ISG MEC}, {Multi-access Edge Computing (MEC); Framework and Reference
  Architecture} (12 2020).

\bibitem{2018_comst_mec_iot}
P.~{Porambage}, J.~{Okwuibe}, M.~{Liyanage}, M.~{Ylianttila}, T.~{Taleb},
  {Survey on Multi-Access Edge Computing for Internet of Things Realization},
  {IEEE Communications Surveys \& Tutorials} 20~(4) (2018) {2961--2991}.
\newblock \href {https://doi.org/10.1109/COMST.2018.2849509}
  {\path{doi:10.1109/COMST.2018.2849509}}.

\bibitem{2021_jnca_edge}
S.~Bolettieri, R.~Bruno, E.~Mingozzi, {Application-aware resource allocation
  and data management for MEC-assisted IoT service providers}, {Journal of
  Network and Computer Applications} 181 (2021) 103020.

\bibitem{NFV028}
ETSI, Network functions virtualisation (nfv) release 3;management and
  orchestration; report on architecture options to support multiple
  administrative domains, Group Report ETSI GR NFV-IFA 028 V3.1.1, ETSI ISG NFV
  (01 2018).

\bibitem{NFV012}
ETSI, Network functions virtualisation (nfv) release 3;evolution and ecosystem;
  report on network slicing support with etsi nfv architecture framework, Group
  Report ETSI GR NFV-EVE 012 V3.1.1, ETSI ISG NFV (01 2018).

\bibitem{NGMN028}
{NGMN}, {Description of Network Slicing Concept}, {White Paper} (Jan. 2016).

\bibitem{3GPPTR28801}
3GPP, {Telecommunication management; Study on management and orchestration of
  network slicing for next generation network (Release 15)}, {Technical Report}
  {28.801 V15.1.0}, {} (Jan. 2018).

\bibitem{IETFFarrel}
A.~Farrel, E.~Gray, J.~Drake, R.~Rokui, S.~Homma, K.~Makhijani, L.~Contreras,
  J.~Tantsura, Framework for ietf network slices, {IETF Draft}
  {draft-ietf-teas-ietf-network-slices-05}, IETF ({October 25} 2021).

\bibitem{ZSM003}
ETSI, Zero-touch network and service management (zsm); end-to-end management
  and orchestration of network slicing, Group Specification ETSI GS ZSM 003
  V1.1.1, ETSI ISG ZSM (06 2021).

\bibitem{MEC024}
{ETSI MEC ISG}, {Multi-access Edge Computing (MEC); Support for network
  slicing}, {Group Report} {MEC 024 V2.1.1}, {ETSI} (Nov. 2019).

\bibitem{Cominardi2020}
L.~Cominardi, T.~Deiss, M.~Filippou, V.~Sciancalepore, F.~Giust, D.~Sabella,
  {MEC Support for Network Slicing: Status and Limitations from a
  Standardization Viewpoint}, {IEEE Communications Standards Magazine} 4~(2)
  (2020) 22--30.
\newblock \href {https://doi.org/10.1109/MCOMSTD.001.1900046}
  {\path{doi:10.1109/MCOMSTD.001.1900046}}.

\bibitem{2020_MNET_MEC_subslice}
A.~Ksentini, P.~A. Frangoudis, {Towards Slicing-Enabled Multi-Access Edge
  Computing in 5G}, IEEE Network 34~(2) (2020) 99--105.

\bibitem{2011_commag_vlan}
M.~Yu, J.~Rexford, X.~Sun, S.~Rao, N.~Feamster, {A survey of virtual LAN usage
  in campus networks}, {IEEE Communications Magazine} 49~(7) (2011) 98--103.

\bibitem{2018_COMST_slicing_survey}
I.~Afolabi, T.~Taleb, K.~Samdanis, A.~Ksentini, H.~Flinck, {Network Slicing and
  Softwarization: A Survey on Principles, Enabling Technologies, and
  Solutions}, {IEEE Communications Surveys \& Tutorials} 20~(3) (2018)
  2429--2453.
\newblock \href {https://doi.org/10.1109/COMST.2018.2815638}
  {\path{doi:10.1109/COMST.2018.2815638}}.

\bibitem{2020_ACCESS_slicing_survey}
L.~U. Khan, I.~Yaqoob, N.~H. Tran, Z.~Han, C.~S. Hong, {Network Slicing: Recent
  Advances, Taxonomy, Requirements, and Open Research Challenges}, {IEEE
  Access} 8 (2020) 36009--36028.

\bibitem{2017_foukas_5G_slicing}
X.~Foukas, G.~Patounas, A.~Elmokashfi, M.~K. Marina, {Network Slicing in 5G:
  Survey and Challenges}, {IEEE Communications Magazine} 55~(5) (2017) 94--100.
\newblock \href {https://doi.org/10.1109/MCOM.2017.1600951}
  {\path{doi:10.1109/MCOM.2017.1600951}}.

\bibitem{3GPPTS28530}
3GPP, Management and orchestration; concepts, use cases and requirements
  (release 17), {Technical Specification} {28.530 V17.1.0}, {} (March 2021).

\bibitem{2021_gsma_TR_e2eslicing}
{GSMA}, {E2E Network Slicing Architecture}, {White Paper}, {} (June 2021).

\bibitem{2017_JCSI_mano}
C.~Rotsos, D.~King, A.~Farshad, J.~Bird, L.~Fawcett, N.~Georgalas, M.~Gunkel,
  K.~Shiomoto, A.~Wang, A.~Mauthe, N.~Race, D.~Hutchison, {Network service
  orchestration standardization: A technology survey}, {Computer Standards \&
  Interfaces} 54 (2017) 203--215.

\bibitem{2020_JSAN_mano}
P.~Trakadas, P.~Karkazis, H.~C. Leligou, T.~Zahariadis, F.~Vicens, A.~Zurita,
  P.~Alemany, T.~Soenen, C.~Parada, J.~Bonnet, E.~Fotopoulou, A.~Zafeiropoulos,
  E.~Kapassa, M.~Touloupou, D.~Kyriazis, {Comparison of Management and
  Orchestration Solutions for the 5G Era}, {Journal of Sensor and Actuator
  Networks} 9~(1) (2020).

\bibitem{2020_COMNET_programmable_5G}
L.~Bonati, M.~Polese, S.~D'Oro, S.~Basagni, T.~Melodia, {Open, Programmable,
  and Virtualized 5G Networks: State-of-the-Art and the Road Ahead}, {Computer
  Networks} 182 (2020) 107516.
\newblock \href {https://doi.org/https://doi.org/10.1016/j.comnet.2020.107516}
  {\path{doi:https://doi.org/10.1016/j.comnet.2020.107516}}.

\bibitem{MEC027}
{ESTI MEC ISG}, {Multi-access Edge Computing(MEC); Study on MEC support for
  alternative virtualization technologies}, {Group Report} {MEC 027 V2.1.1},
  {ETSI} (Nov. 2019).

\bibitem{2021_TNSM_e22_slice_survey}
M.~Chahbar, G.~Diaz, A.~Dandoush, C.~C{\'e}rin, K.~Ghoumid, {A Comprehensive
  Survey on the E2E 5G Network Slicing Model}, {IEEE Transactions on Network
  and Service Management} 18~(1) (2021) 49--62.

\bibitem{3GPPTS28531}
3rd Generation Partnership~Project, Management and orchestration; provisioning;
  (release 17), Technical Specification 3GPP TS 28.531 V17.0.0, 3GPP (06 2021).

\bibitem{3GPPTS28541}
3rd Generation Partnership~Project, Management and orchestration; 5g network
  resource model (nrm); stage 2 and stage 3 (release 17 ), Technical
  Specification 3GPP TS 28.541 V17.3.0, 3GPP (06 2021).

\bibitem{2020_broadnets_mec_k8s}
I.~D. Mart{\'\i}nez-Casanueva, L.~Bellido, C.~M. Lentisco, D.~Fern{\'a}ndez,
  {An Initial Approach to a Multi-access Edge Computing Reference Architecture
  Implementation Using Kubernetes}, in: {EAI International Conference on
  Broadband Communications, Networks, and Systems (BROADNETS)}, {Qingdao,
  China}, 2020, pp. {1--9}.

\bibitem{NFV-SOL005}
ETSI, {Network Functions Virtualisation (NFV) Release 2; Protocols and Data
  Models; RESTful protocols specification for the Os-Ma-nfvo Reference Point},
  Group Specification ETSI GS NFV-SOL 005 V2.7.1, {ETSI ISG NFV} (January
  2020).

\bibitem{2019_netsoft_netaware_kube}
J.~Santos, T.~Wauters, B.~Volckaert, F.~De~Turck, {Towards Network-Aware
  Resource Provisioning in Kubernetes for Fog Computing Applications}, in:
  {Proc. of the 5th IEEE Conference on Network Softwarization (NetSoft)}, 2019,
  pp. 351--359.

\bibitem{2020_noms_delayaware_kube}
J.~Santos, T.~Wauters, B.~Volckaert, F.~De~Turck, {Towards delay-aware
  container-based Service Function Chaining in Fog Computing}, in: {Proc. of
  IEEE/IFIP Network Operations and Management Symposium (NOMS)}, 2020, pp.
  1--9.
\newblock \href {https://doi.org/10.1109/NOMS47738.2020.9110376}
  {\path{doi:10.1109/NOMS47738.2020.9110376}}.

\bibitem{2016_bwguar_openflow}
H.~Krishna, N.~L.~M. van Adrichem, F.~A. Kuipers, {Providing bandwidth
  guarantees with OpenFlow}, in: {Proc. of the IEEE Symposium on Communications
  and Vehicular Technologies (SCVT)}, 2016, pp. 1--6.
\newblock \href {https://doi.org/10.1109/SCVT.2016.7797664}
  {\path{doi:10.1109/SCVT.2016.7797664}}.

\bibitem{2018_MCC_TOSCA}
P.~Lipton, D.~Palma, M.~Rutkowski, D.~A. Tamburri, {TOSCA Solves Big Problems
  in the Cloud and Beyond!}, {IEEE Cloud Computing} 5~(02) (2018) 37--47.

\bibitem{NFV-SOL001}
ETSI, {Network Functions Virtualisation (NFV) Release 2; Protocols and Data
  Models; NFV descriptors based on TOSCA specification}, Group Specification
  ETSI GS NFV-SOL 001 V2.7.1, ETSI ISG NFV (May 2021).

\bibitem{2021_jgc_torch}
O.~Tomarchio, D.~Calcaterra, G.~D. Modica, P.~Mazzaglia, {TORCH: a TOSCA-Based
  Orchestrator of Multi-Cloud Containerised Applications}, {Journal of Grid
  Computing} 19 (2021).

\bibitem{2018_icds_edge_share}
T.~{He}, H.~{Khamfroush}, S.~{Wang}, T.~{La Porta}, S.~{Stein}, {It's Hard to
  Share: Joint Service Placement and Request Scheduling in Edge Clouds with
  Sharable and Non-Sharable Resources}, in: {Proc. of IEEE ICDCS'18}, 2018, pp.
  365--375.

\bibitem{2020_mobihoc_si-edge}
S.~D'Oro, L.~Bonati, F.~Restuccia, M.~Polese, M.~Zorzi, T.~Melodia, {Sl-Edge:
  Network Slicing at the Edge}, in: {Proc. of ACM Mobihoc'20}, 2020.

\end{thebibliography}





\end{document}